\documentclass[%
 reprint,
nofootinbib,
 amsmath,amssymb,
 aps,
 onecolumn,
]{revtex4}

\usepackage{amssymb,amsmath,amsfonts,amsbsy,graphicx}
\usepackage{color}
\usepackage{psfrag}
\usepackage{stackrel}

\newcommand\be{\begin{equation}}
\newcommand\ee{\end{equation}}
\newcommand\ba{\begin{eqnarray}}
\newcommand\ea{\end{eqnarray}}\newcommand\eq{\begin{equation}}           
\newcommand\en{\end{equation}}

 \newcount\colveccount
\newcommand*\colvec[1]{
        \global\colveccount#1
        \begin{pmatrix}
        \colvecnext
}
\def\colvecnext#1{
        #1
        \global\advance\colveccount-1
        \ifnum\colveccount>0
                \\
                \expandafter\colvecnext
        \else
                \end{pmatrix}
        \fi
}
\input epsf

\usepackage{ulem}
\usepackage{amssymb}
\usepackage{amsmath}
\usepackage{bm}
\usepackage{graphicx}
\usepackage{color}
\usepackage{subfig}
\usepackage{caption}
\usepackage{url}
\usepackage{amsbsy}

\def\gsim{\;\rlap{\lower 2.5pt
 \hbox{$\sim$}}\raise 1.5pt\hbox{$>$}\;}
\def\lsim{\;\rlap{\lower 2.5pt
 \hbox{$\sim$}}\raise 1.5pt\hbox{$<$}\;}

\usepackage{ragged2e}
\DeclareCaptionJustification{justified}{\justifying}
\begin{document}
\title{
CMB and 21cm bounds on early structure formation boosted by primordial black hole entropy fluctuations
}
\author{Hiroyuki Tashiro$^1$ and Kenji Kadota$^2$ \\
  {\small $^1$ Department of physics and astrophysics, Nagoya University, Nagoya 464-8602, Japan}\\
    {\small $^2$ Center for Theoretical Physics of the Universe, Institute for Basic Science (IBS), Daejeon, 34051, Korea} 
  }
 

\begin{abstract}
The dark matter (DM) can consist of the primordial black holes (PBHs) in addition to the conventional weakly interacting massive particles (WIMPs). The Poisson fluctuations of the PBH number density produce the isocurvature perturbations which can dominate the matter power spectrum at small scales and enhance the early structure formation.  We study how the WIMP annihilation from those early formed structures can affect the CMB (in particular the E-mode polarization anisotropies and $y$-type spectral distortions) and global 21cm signals.
Our studies would be of particular interest for the light (sub-GeV) WIMP scenarios which have been less explored compared with the mixed DM scenarios consisting of PBHs and heavy ($\gtrsim 1$ GeV) WIMPs. For instance, for the self-annihilating DM mass $m_{\chi}=1$~MeV and the thermally averaged annihilation cross section $\langle \sigma v \rangle \sim 10^{-30} \rm  cm^3/s$, the latest Planck CMB data requires the PBH fraction with respect to the whole DM to be at most ${\cal O}(10^{-3})$ for the sub-solar mass PBHs and an even tighter bound (by a factor $\sim 5$) can be obtained from the global 21-cm measurements.
\end{abstract}

\maketitle   

\setcounter{footnote}{0} 
\setcounter{page}{1}\setcounter{section}{0} \setcounter{subsection}{0}
\setcounter{subsubsection}{0}

\section{introduction}
While the properties of the dark matter (DM) still remain unknown, there has been a revived interest in the primordial black hole~(PBH) DM in view of the advancement of gravitational wave experiments \cite{Abbott:2016blz}. While the allowed parameter range for the PBH to be the dominant DM component has been narrowed down, the PBH being a partial component of DM still remains an intriguing possibility \cite{Carr:2020xqk,Green:2020jor,Carr:2020gox}.

We study the effects of the DM annihilation on the cosmic microwave background~(CMB) and 21-cm signals in the mixed DM scenarios consisting of the self-annihilating DM and PBH. 
We focus on the boosted DM annihilation due to the enhanced early structure formation in the presence of the Poisson noise sourced by the PBHs~\cite{Afshordi:2003zb}. 

The randomly distributed PBHs can add the Poisson noise to the matter power spectrum which can dominate the conventional adiabatic perturbations at small scales. 
The possibility for the PBHs to be (partial) DM and the consequent early structure formation due to such PBH sourced isocurvature perturbations have been investigated to seek the potential signals of the PBHs, and the astrophysical probes sensitive to those enhanced small scale structures such as the gravitational lensing and Ly-$\alpha$ observations have been explored~\cite{Afshordi:2003zb,Ali-Haimoud:2018dau,Inman:2019wvr,Desjacques:2018wuu,Oguri:2020ldf,Kadota:2020ahr,Gong:2017sie,Mena:2019nhm}. 

When there exists self-annihilating DM (a typical example is WIMP), in addition to PBH as a partial DM, the thermal evolution of baryons in the dark ages is also affected due to the enhanced DM annihilation from those abundant early formed halos.  
The additional energy injection from DM annihilation may produce observable CMB distortions~\cite{1969Ap&SS...4..301Z,1970Ap&SS...7...20S,1991A&A...246...49B,Hu:1993gc}.
The early reionization due to the enhancement of DM annihilation affects the Thomson optical depth and modifies the CMB angular power spectrum~\cite{Aghanim:2019ame}. 
The measurement of global 21-cm line signal is also expected to be a useful tool to probe the thermal history of the Universe~\cite{1997ApJ...475..429M,1999A&A...345..380S,2010PhRvD..82b3006P}.
The recent result reported by the EDGES~\cite{2018Natur.555...67B} motivates to study the constraint on extra heat sources~\cite{Clark2018,Hektor2018a,Mitridate2018,Cheng2018,DAmico2018,Safarzadeh2018,Minoda:2018gxj}
and the nature of DM~\cite{Tashiro:2014tsa,Barkana:2018lgd}.
We in this paper quantify the effects of the boosted DM annihilation on the CMB anisotropy (in particular the E-mode polarization power spectrum), CMB spectral distortion and global 21cm signals.

The early structure formation can result in abundant small halos some of which can potentially survive the tidal disruptions due to the tightly bound structures. We calculate the effects of DM annihilation from the substructures due to survived small halos as well as those from the halos without considering the substructures. 
For instance, for $m_{\chi}=1$MeV and $\langle \sigma v \rangle \sim  10^{-30}~\rm cm^2$, the Planck CMB data can give the bound on the PBH fraction to the total DM $f_{\rm PBH}\lesssim 10^{-2.7}$ for the sub-solar mass PBH assuming the survival of the halos if produced at $z>250$ (note the CMB bound in the absence of PBHs is $\langle \sigma v \rangle/m_{\chi}\lesssim 3 \times 10^{-27} \rm cm^3/s/GeV$ for the DM with s-wave annihilation \cite{Slatyer:2015jla}). We for simplicity assume the monochromatic mass function for the PBHs throughout this paper. We also find that the effects of the survived minihalos are negligible for $f_{\rm PBH}\lesssim 10^{-2.6}$, and the aforementioned bound $f_{\rm PBH}\lesssim 10^{-2.7}$ hence can be considered as the robust bound arising only from isolated halos without including the effects of substructures. We also show that, using the redshift dependence of the EDGES data (without using their larger-than-expected absorption signal amplitude information), the 21cm global signal can lead to the tighter bounds on $f_{\rm PBH}$ by a factor 5 than that from the CMB.
 
We mention that, independently from such 'Poisson effects' (PBHs' collective effects on the large scale structure), one can also study the 'seed effects' (the DM accretion into an individual PBH) which can be complimentary to the studies in this paper. 
The mixed DM scenarios of PBHs and self-annihilating WIMPs indeed have been actively investigated for the scenarios where the PBH can be a 'seed' for the ultracompact minihalo~(UCMH) with a steep DM density profile \cite{Gondolo:1999ef,Lacki:2010zf,Boucenna:2017ghj,Adamek:2019gns,Eroshenko:2016yve,Carr:2020mqm,Cai:2020fnq,Delos:2018ueo,Kohri:2014lza,Bertone:2019vsk,Carr:2018rid,1984MNRAS.206..801C,Hertzberg:2020kpm,Kashlinsky:2016sdv,Tashiro:2021xnj}. No observation of possible enhanced DM annihilation signals from such a DM density spike around a PBH can lead to the severe bounds on the allowed abundance of PBHs, and the incompatibility of co-existence of the PBH and WIMP has been pointed out.  
Such 'seed effects' have been mainly discussed for the heavy ($>1$ GeV) DM and the conventional cross sections of order $\langle \sigma v \rangle \sim {\cal O}(10^{-26})~\rm cm^3/s$. For instance, the tight bounds $f_{\rm PBH}\lesssim {\cal O}(10^{-9})$ can be obtained for the typical WIMP parameters (e.g. $m_{\chi}\sim 100~{\rm GeV}, \langle \sigma v \rangle=3\times 10^{-26}~{\rm cm^3/s}$) due to no detection of such enhanced DM annihilation signals in the gamma ray or CMB data \cite{Adamek:2019gns,Tashiro:2021xnj}. Those tight bounds on $f_{\rm PBH}$ are usually derived based on the characteristic steep DM density profile $\rho \propto r^{-\gamma}$ (e.g. $\gamma\sim 9/4$), and those bounds cannot be straightforwardly applied when the DM kinetic energy cannot be ignored compared with its potential energy in estimating the halo profile around a PBH \cite{Gondolo:1999ef,Lacki:2010zf,Boucenna:2017ghj,Adamek:2019gns,Eroshenko:2016yve,Carr:2020mqm,Cai:2020fnq,Delos:2018ueo,Kohri:2014lza,Bertone:2019vsk}. Such a UCMH profile in the presence of the PBH for the non-trivial initial conditions including those for the light (e.g. $\lesssim 1$ MeV) DM still has not been fully explored by the numerical simulations yet without definite answers for the bounds on the PBH parameters. Fortunately, our Poisson bounds are independent from those bounds involving the non-trivial DM accretion to the PBHs and hence can offer complementary bounds to those seed effects. Our studies would be of particular interest for the light (sub-GeV) DM for which the UCMH profile is less steep and the seed effect bounds are less severe. The bounds from the DM accretion onto the PBH are also relaxed for the lighter PBH mass as well as for the lighter DM mass, and we focus our discussions on the light (sub-GeV) DM mass and the light PBH mass (sub-solar mass) even though our study can be straightforwardly extended to the larger DM and PBH masses for which the PBH seed effects however would give more stringent bounds on $f_{\rm PBH}$ than the Poisson effects.

Our paper is organized as follows. Sec. \ref{sec2} reviews the matter power spectrum in the presence of the PBH isocurvature perturbations and specifies the PBH parameter range to be explored in our study. Sec. \ref{sec3} outlines the formalism to calculate the DM annihilation boost in the presence of the early formed structures. Sec \ref{sec4} discusses how the evolutions of ionization fraction and baryon temperature are affected due to the DM annihilation in the presence of the PBHs. Such changes in the early history of the Universe are implemented for the calculations of the bounds on $f_{\rm PBH}$ from the CMB (fluctuation anisotropy and spectral distortion) and 21cm signals in Sec \ref{sec5}.

\section{The Poisson fluctuations due to PBHs}
\label{sec2}
\begin{figure}[htbp]
     \begin{tabular}{c}

                           \includegraphics[width=0.5\textwidth]{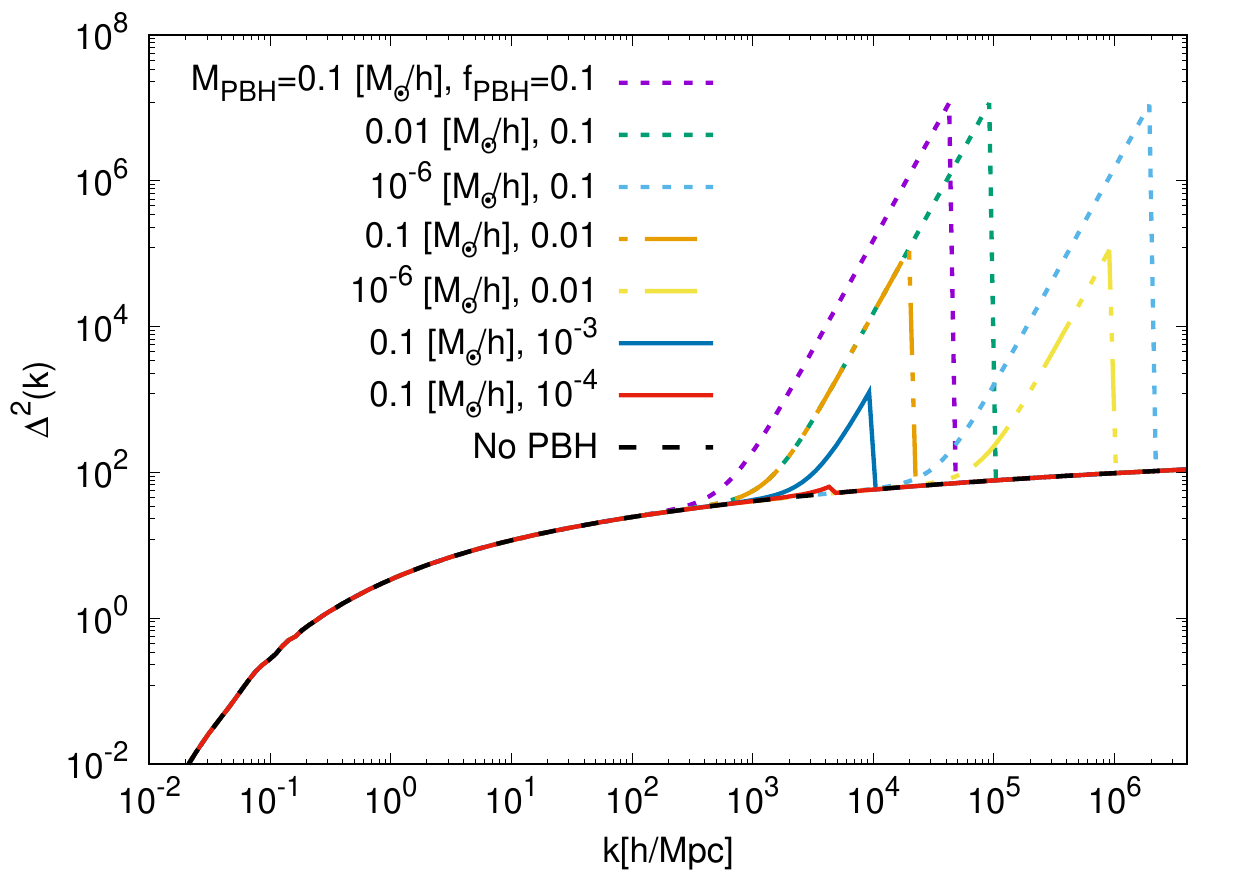} 

    \end{tabular}

    
   \caption{
     The linear matter power spectrum in the mixed DM scenarios at $z=0$. The PBH contributions are parameterized by its fraction and mass $f_{\rm PBH}\equiv \Omega_{\rm PBH}/\Omega_{\rm DM}, M_{\rm PBH}$. The scenario with the PBH fraction smaller than $10^{-4}$ is not distinguishable from the standard $\Lambda$CDM with no PBH in this figure. 
   }
%
   \label{kvspk}
\end{figure}

 The PBHs are assumed to be randomly distributed\footnote{This is a reasonable assumption because  the typical separation between the PBHs which can be formed before the matter-radiation equality epoch would be larger than the horizon scale at the formation of PBHs.}, and the PBHs can contribute to the matter power spectrum as the Poisson noise 
\ba
P_{\rm PBH}=\frac{1}{n_{\rm PBH}}, ~\quad n_{\rm PBH}=\frac{\Omega_{\rm DM}\rho_{\rm cri}f_{\rm PBH}}{M_{\rm PBH}}, 
\ea
where $f_{\rm PBH}\equiv \Omega_{\rm PBH}/\Omega_{\rm DM}$ represents the fraction of PBH contribution to the total DM. We focus on the scales larger than the mean separation of the PBHs, $k\lesssim k_*= n_{\rm PBH}^{1/3}$, so that we treat the PBHs as the ideal pressureless fluid. Such PBH fluctuations appear only in the PBH component and independent from the adiabatic perturbations, and the PBH Poisson fluctuations can be treated as the isocurvature contribution to the total power spectrum  \cite{Afshordi:2003zb,Ali-Haimoud:2018dau,Inman:2019wvr,Desjacques:2018wuu}
\ba
P(k,z)=D^2(z) \left(
T^2_{ad}(k) P_{\rm adi}(k)+ T^2_{\rm iso}P_{\rm iso}(k)
\right),
\ea
where $P_{\rm adi}$ is the conventional adiabatic power spectrum and $P_{\rm iso}=f_{\rm PBH}^2P_{\rm PBH}$. $D(z)$ is the growth function normalized by $D(0)=1$ and $T$ represents the transfer function. The isocurvature transfer function $T_{\rm iso}$ reads \cite{1999coph.book.....P}
\ba
T_{\rm iso}(k)&=&\frac{3}{2} (1+z_{\rm eq}) \quad \mbox{ for }k_{\rm eq}<k<k_*,\\
T_{\rm iso}(k)&=&0 \quad \mbox{ otherwise.}
\ea
The dimensionless power spectra $\Delta^2(k)=P(k)k^3/2\pi^2 $ in the presence of PBH are shown in Fig.~\ref{kvspk}. 
The dimensionless power spectrum $\Delta^2(k)$ is smaller with a smaller $M_{\rm PBH}$ for a given $k$, but the cutoff $k_*$ is bigger for a smaller $M_{\rm PBH}$ leading to the common peak height with a different $M_{\rm PBH}$ for a given $f_{\rm PBH}$.
   \begin{figure}[htbp]
          \includegraphics[width=0.5\textwidth]{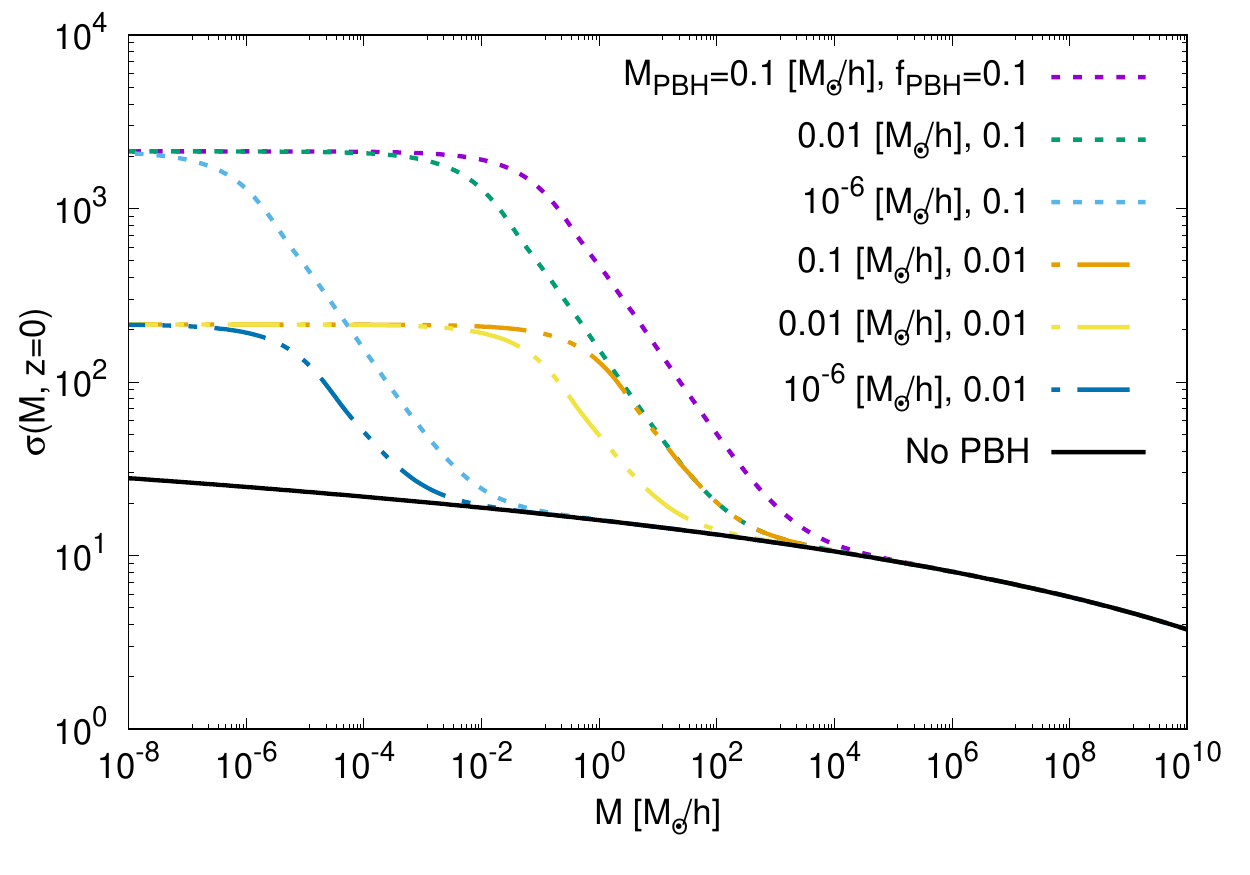}
     
   \begin{tabular}{c}
    \end{tabular}

    
   \caption{
The mass variance $\sigma(M, z=0)$ in the mixed DM scenarios. The PBH mass dependence becomes small for a small halo mass scale $M \lesssim 10^{-6} M_{\odot}$ in the parameter range of our interest $f_{\rm PBH}\lesssim 0.1$, $10^{-6} [M_{\odot}/h] \lesssim M_{\rm PBH} \lesssim 1 [M_{\odot}/h]$.  
   }

%
   
   \label{sigmaz0}
\end{figure}
This independence of the peak amplitudes on $M_{\rm PBH}$ is also reflected in the insensitivity of $\sigma$ to $M_{\rm PBH}$ for a small halo mass as illustrated in Fig. \ref{sigmaz0}. The variance of the mass fluctuations smoothed on the scale of a sphere containing a mass $M$ is defined by
\ba
\sigma^2(M,z)=\int d\ln k \frac{k^3 P(k,z)}{2\pi^2} \left| W(kR)\right|^2,
\ea
where $W(kR)=3\left[\sin(kR)-(kR)\cos(kR)\right]/(kR)^3$ is the Fourier transform of the real-space spherical top-hat window function which contains mass $M$. 
 
 We study the effects of early formed minihalos on the CMB and 21cm signals, and the parameters, besides the conventional $\Lambda$CDM parameters, which are relevant in our study are the minimal halo mass $M_{\rm min}$, PBH fraction and mass~($f_{\rm PBH}$ and $M_{\rm PBH}$) and the annihilating DM mass and the thermally averaged DM annihilation cross section~($m_{\chi}$ and $\langle \sigma v\rangle$).
The minimum halo mass can vary in a wide range (for instance $10^{-4}M_{\odot}/h$ to $10^{-12}M_{\odot}/h$) which is heavily dependent on the nature of DM kinetic decoupling affecting the DM free streaming and acoustic damping. We for concreteness use the conventional value of $M_{\rm min}=10^{-6} M_{\odot}/h$ in our calculations \cite{Bertschinger:2006nq,Loeb:2005pm,Gondolo:2012vh,Profumo:2006bv,Gondolo:2016mrz,Green:2003un,Green:2005fa,Bringmann:2006mu,Diemand:2005vz}. 
The mass scale of $10^{-6} M_{\odot}/h$ corresponds to the comoving scale of order $k\sim 10^6 h/\rm Mpc$, and we focus on the PBH parameters whose Poisson noise cutoff $k_* < 10^6 h/\rm Mpc$ so that the PBH fluctuation cutoff shows up above the DM minimum halos mass scale. For instance, $k_* \gtrsim 4 \times 10^6 h/\rm Mpc$ for $M_{\rm PBH}\lesssim 10^{-7} M_{\odot}/h$ with $f_{\rm PBH}=0.1$. 
We hence for concreteness focus on $M_{\rm PBH}>10^{-6}M_{\odot}$ in our discussions unless stated otherwise. We also note, for this parameter range of our interest, $\sigma(M_{\rm min})$ is not dependent on $M_{\rm PBH}$ for a given $f_{PBH}$. 
Due to this insensitivity of $\sigma(M_{\rm min})$ on $M_{\rm PBH}$, our constraints on $f_{\rm PBH}$ turn out to be insensitive to $M_{\rm PBH}$ as discussed in the following analysis. 

\section{Annihilation boosting by the early structure formation due to PBHs}
\label{sec3}
The annihilation of self-annihilating DM is proportional to the density squared and the ratio of the fluctuation contribution to the smooth background contribution is the so-called boost factor $B$
\ba
\langle \rho_{\rm DM}^2(z) \rangle
= \bar{\rho}_{\rm DM}^2(z) \langle  (1+\delta(z)) ^2 \rangle=
\bar{\rho}_{\rm DM}^2(z) (1+\langle \delta^2(z)\rangle)
\equiv  \bar{\rho}_{\rm DM}^2(z) (1+B(z))
\ea
where $\bar{\rho}_{\rm DM}$ is the homogeneous DM background density and $\delta=(\rho - \bar{\rho})/\bar{\rho}$ is the density contrast.
We adopt the halo model approach in calculating the boost factor to take account of the DM structure formation. The energy injection rate per volume due to the DM annihilation (we assume the self-annihilating Majorana DM $\chi$) into the cosmic plasma reads \cite{Cooray:2002dia, Poulin:2015pna}
\be
\label{eq:dEdVdt}
 \frac{d^2E}{dVdt} (z) = \left. \frac{d^2E}{dVdt} \right\vert_{\rm sm}+\left. \frac{d^2E}{dVdt} \right\vert_{\rm str},
\ee 
where the subscript sm denotes the smooth background contributions and str represents the DM halo contributions to account for the effects from the structure formation.

The smooth background contribution is obtained from
\be
\label{eq:dEdVdt_smooth}
\left. \frac{d^2E}{dVdt} \right\vert_{\rm sm}= f_{\rm ann }\frac{\langle \sigma v \rangle}{ m_{\chi}}\rho_{\rm cri}^2 \Omega^2_{\chi}(1+z)^6,
\ee
where $f_{\rm ann }$ is the fraction of the energy absorbed into the plasma at a redshift $z$.
Although it depends on the annihilating DM mass and the redshift, we assume for simplicity that $f_{\rm ann }$ is constant ~\cite{Galli:2013dna}. $\Omega_\chi$ is the energy density parameter of the self-annihilating DM $\chi$. Since we also consider PBHs as DM, 
$\Omega_{\rm DM} = \Omega_{\rm PBH} + \Omega_\chi$.

The halo contribution is given as~\cite{Giesen:2012rp}
\be
\label{eq:ordinary_halo}
\left. \frac{d^2E}{dVdt} \right\vert_{\rm halo} =
 f_{\rm ann }\frac{\langle \sigma v \rangle}{ m_{\chi}} (1+z)^3  \int^{\infty}_{M_{\rm min}}dM~
\frac{dn}{dM}(M,z)
\left(
\int ^{r_{200}}_0
dr~
4\pi r^2 \rho_{\rm halo}^2(r)
\right),
\ee
where 
$M_{\rm min}$ is the minimum mass of DM halos,
$dn/dM$ is the mass function of DM halos and $r_{200}$ is the typical radius of a DM halo with mass $M$ in which an averaged density equals to 200 times the background density and $\rho_{\rm halo}$ is the radial density profile of self-annihilating DM.

When we use the NFW profile for the halo density $\rho_{\rm halo}$ and take into account that only a fraction $\Omega_{\chi}/\Omega_{\rm M}$ of the total matter can annihilate,
the integral of the self-annihilating DM density profile can be written as
\ba
&&\int ^{r_{200}}_0
dr
4\pi r^2 \rho_{\rm halo}^2(r)
=\frac{M \bar{\rho}(z_{\rm f})}{3}
\left(
\frac{\Omega_{\chi}}{\Omega_{\rm M}}
\right)^2,
\\
&&\bar{\rho}(z_{\rm f})=200 \rho_{\rm cri} \Omega_{\rm M} (1+z_{\rm f})^3 F(c(z_{\rm f})), ~\quad
F(c(z_{\rm f}))=\frac{c^3}{3} \frac{1-(1+c)^{-3}}{\left( \ln(1+c)-c(1+c)^{-1}  \right)^2},
\ea 
where $z_{\rm f}$ is the formation redshift of DM halos. $c(z_{\rm f})$ is the concentration parameter of the NFW profile for DM halos forming at $z_{\rm f}$.
We conservatively take $c(z_{\rm f})=2$ for all the early formed halos because it is the minimal value of the concentration parameters found in the simulations performed for a wide range of halo masses \cite{Wang:2019ftp,Ishiyama:2020vao,Sanchez-Conde:2013yxa,Moline:2016pbm,Zhao:2008wd,Prada:2011jf,Ackermann:2015tah}. A bigger concentration parameter results in a bigger annihilation rate.

Assuming the Press-Schechter (PS) mass function can simplify the integral of the mass function and the mass fraction in collapsed objects, $f_{\rm coll}$, becomes, according to the PS formalism,
\be
\label{fcolleq}
f_{\rm coll}(z) = \int^{\infty}_{M_{\rm min}}dM~ \frac{M}{\Omega_{\rm M} \rho_{\rm cri}}
\frac{dn}{dM}(M,z)= 
{\rm erfc}(u_{\rm min}(z)), \quad u_{\rm min}(z)=\frac{\delta_c}{\sqrt2 \sigma( M_{\rm min},z)} ,
\ee 
where $\delta_{\rm c}\approx 1.686$ is the critical density contrast for collapse. 
Therefore, we can rewrite the DM halo contribution as
\be
\label{eq:halo_fcoll}
\left. \frac{d^2E}{dVdt} \right\vert_{\rm halo} =
\frac{\bar{\rho}(z_{\rm f})}{3}
\left(
\frac{\Omega_{\chi}}{\Omega_{\rm M}}
\right)^2
f_{\rm ann } \frac{\langle \sigma v \rangle}{ m_{\chi}} (1+z)^3 
 f_{\rm coll}(z).
\ee
which represents the energy injection at the redshift $z$ from the halos formed at $z_{\rm f}$. 



In a simple treatment where the substructure contribution is ignored (such as in the conventional PS formalism which does not take account of subhalo abundance once they merge into the larger halos), $z_{\rm f}$ is set to $z_{\rm f} = z$ in Eq.~\eqref{eq:halo_fcoll}.
The notable feature of our PBH scenarios arises due to the enhancement of early structure formation caused by the PBH isocurvature fluctuations. Those early formed halos are shown to be dense enough (the density $\propto (1+z_{\rm f})^3$) to survive the potential tidal disruptions for $z_{\rm f}\gtrsim 250$ \cite{Blinov:2019jqc,Kadota:2020ahr}\footnote{The tidal disruptions include the stellar encounters, encountering among small halos and tidal stripping at the core of a host halo after the infall due to the dynamical friction. We refer the readers to Refs.~\cite{Zhao:2005mb,vandenBosch:2017ynq,Green:2019zkz,Kadota:2020ahr,Blinov:2019jqc,Berezinsky:2003vn,Arvanitaki:2019rax,Berezinsky:2005py,Berezinsky:2003vn,2020AJ....159...49D,Xiao:2021nkb} for the further discussions on the survival of the early formed halos. One can analytically estimate the time-scale of tidal disruptions due to stellar encounters (which are the most significant tidal disruption processes), based on the impulse approximation, as \cite{Kadota:2020ahr}
\ba
t \sim 70 Gyr \left( \frac{1+z_f}{100} \right)^{3/2} \left( \frac{10^6 M_{\odot}/kpc^3}{\bar{n}_* M_*} \right) 
\ea
where $M_*$ is the mass of a stellar object (such as a star) and $\bar{{n}}_*$ is the corresponding mean number density of stars in a host halo.
We can hence infer that the minihalos in a galactic host halo could well survive if they were formed at $z_f\geq 100$.} Such early formed dense structures resilient to the tidal disruptions make our scenarios strikingly different from the $\Lambda$CDM for which the typical halo formation occurs at $z\lesssim 20$.

We introduce a characteristic redshift $z_*$ (for concreteness, we use $z_*=250$ in our calculations \cite{Blinov:2019jqc,Kadota:2020ahr}). We call the small halos produced at $z_{\rm f}>z_*$ 'minihalos' and assume those minihalos can survive until now while keeping the density $200 \bar \rho(z_{\rm f})$ at the forming epoch for simplicity. On the other hand, we assume the halos forming at $z_{\rm f} < z_*$ cannot survive and lose their identities in the hierarchical structure formation. In other words, for the halos formed at $z_{\rm f}<z_*$, we apply the conventional PS formalism (we hence do not account for the substructure contribution and ignore their survivals at the later epoch as minihalos). The density of such halos at a redshift $z$ is 
simply proportional to the background density $200  \bar \rho(z)$.
The actual DM annihilation effects could be bigger than our estimations for those halos produced at $z<z_*$ because we completely ignore their survival. For the minihalos collapsed at $z>z_*$, we assume the DM annihilation continues in those dense surviving minihalos even after their collapse epochs. 
The effects of the survived minihalos compared with those produced at $z<z_*$ become more prominent at a lower redshift because of a bigger relative difference in $(1+z)^3$ factor in their densities. 
This can be seen, for instance, in Fig. \ref{fig:xe_temp} where the baryon temperature in the presence of the survived minihalos exceeds that without substructures at a low redshift $z\lesssim 50$.

In order to take into account those early formed minihalos,
we divide the contribution from the DM structure formation in Eq.~\eqref{eq:dEdVdt}
into two parts,
\be
\label{eq:dEdVdt_halo}
\left. \frac{d^2E}{dVdt} \right\vert_{\rm str} =
\left. \frac{d^2E}{dVdt} \right\vert_{\rm mh}+\left. \frac{d^2E}{dVdt} \right\vert_{\rm halo}.
\ee 
Here the term with the subscript mh represents the contribution from the survived minihalos forming before~$z_*$. 
Analogously to Eq.~\eqref{eq:ordinary_halo},
the minihalo contribution at a redshift~$z$ can be given as
\be
\label{eq:sub_halo}
\left. \frac{d^2E}{dVdt} \right\vert_{\rm mh} =f_{\rm ann }
 \frac{\langle \sigma v \rangle}{ m_{\chi}} (1+z)^3 
 \int_{z_{\rm min}} ^{z_{\rm max}} dz_{\rm f} 
 \int^{\infty}_{M_{\rm min}}dM~
\frac{d}{dz_{\rm f}} \left[
\frac{dn_{\rm mh}}{dM}(M,z_{\rm f})
\right]
\left(
\int ^{r_{200}}_0
dr~
4\pi r^2 \rho_{\rm halo}^2(r)
\right),
\ee
where ${dn_{\rm mh}}/{dM}(M,z_{\rm f})$ is the mass function of minihalos forming at $z_{\rm f}$.
In the above equation, $z_{\rm min} = {\rm max}[z_*, z]$.
The redshift~$z_{\rm max}$ represents the maximum redshift for the structure formation. 
Although the DM density fluctuation can grow logarithmically in the radiation dominated epoch, it cannot collapse until the matter dominated epoch. 
In our fiducial model, we set $z_{\rm max} = 2000$ and we conservatively assume that, even if its density contrast exceeds the critical density contrast for collapse before $z_{\rm max}$, the overdensity region collapses to a minihalo at $z_{\rm max}$.

In estimating Eq.~\eqref{eq:sub_halo}, assuming that DM halos forming before $z_*=250$ can survive as minihalos, we can consider the fraction of matter collapsed into minihalos $f_{\rm coll,mh}$ as
\be
\label{eq:fcoll_sub}
f_{\rm coll,mh}(z) =  f_{\rm coll} ({\rm max}[z_*, z]),
\ee 
which can also be written as
\be
f_{\rm coll,mh}(z) =
\int_z^{z_{\rm max}} dz_{\rm f}~\frac{df_{\rm coll,mh}}{dz_{\rm f}}(z,z_{\rm f})
=\frac{1}{\rho_{\rm cri} \Omega_{\rm M}} \int_z dz_{\rm f}~
\int_{M_{\rm min}}^{\infty} dM M \frac{d}{dz_{\rm f}} \left[
\frac{dn_{\rm mh}}{dM}(M,z_{\rm f})
\right].
\ee
Note that our formalism does not require the explicit expression for differential fraction $df_{\rm coll,mh}/dz_{\rm f}$ or minihalo mass function $dn_{\rm mh}/dM (M,z_{\rm f})$ because $f_{\rm coll,mh}$ can be obtained from Eqs. \eqref{fcolleq} and \eqref{eq:fcoll_sub} (see Ref. \cite{Delos:2018ueo} for more details on the derivation of differential mass fraction $df/dz$). 
Accordingly, the minihalo contribution can be rewritten with $f_{\rm coll,mh}$ 
\be
\left. \frac{d^2E}{dVdt} \right\vert_{\rm mh} =
\left(
\frac{\Omega_{\chi}}{\Omega_{\rm M}}
\right)^2
 f_{\rm ann }\frac{\langle \sigma v \rangle}{ m_{\chi}} (1+z)^3 
\int dz_{\rm f}
\frac{\bar{\rho}(z_{\rm f})}{3}
\frac{d f_{\rm coll,mh}}{dz_{\rm f}}
(z,z_{\rm f}).
\label{eq:dEdVdt_sub}
\ee

In Eq.~\eqref{eq:dEdVdt_halo}, ${d^2E}/{dVdt} |_{\rm halo}$
represents the contribution from DM which is not included in a survived minihalo formed at $z>z_*$ but  resides in a halo formed at $z<z_*$.
The collapse fraction of such DM is given by
$f_{\rm coll,halo} = f_{\rm coll} - f_{\rm coll,mh}$.
Avoiding the double counting of DM in minihalos in Eq.~\eqref{eq:halo_fcoll}, ${d^2E}/{dVdt} |_{\rm halo}$ is given as
\be
\label{eq:dEdVdt_main}
\left. \frac{d^2E}{dVdt} \right\vert_{\rm halo} =
\frac{\bar{\rho}(z_{\rm f})}{3}
\left(
\frac{\Omega_{\chi}}{\Omega_{\rm M}}
\right)^2 f_{\rm ann }
 \frac{\langle \sigma v \rangle}{ m_{\chi}} (1+z)^3 
 f_{\rm coll,halo}(z).
\ee

The boost factor also consists of two parts,
\be
B(z)=B_{\rm mh}(z)+B_{\rm halo}(z).
\ee

Each boost factor is given by the ratio of Eq.~(\ref{eq:dEdVdt_sub}, \ref{eq:dEdVdt_main})
to the smooth background contribution, Eq.~\eqref{eq:dEdVdt_smooth}.
Therefore, we can calculate boost factors from

\ba
B_{\rm mh}(z)&=&\frac{
200 
}{3}
 \int^{z_{\rm mh}}
 _{z} d z_{\rm f}
  \frac{(1+z_{\rm f})^3}{(1+z)^3} F(c(z_{\rm f}))
 \frac{df_{\rm coll,mh}}{dz_{\rm f}} (z,z_{\rm f}) , 
 \\
 B_{\rm halo}(z)&=&
\frac{
200 
}{3}F(c(z))
f_{\rm coll,halo}(z).
\ea

\section{Evolution of the ionization fraction and the baryon temperature}
\label{sec4}


The injected energy due to the WIMP annihilation can affect the evolutions of the ionization fraction and temperature in the cosmic plasma.
Including the WIMP annihilation effects, these evolutions are calculated from 
\begin{align}
(1+z)\frac{d x_e}{d z} &= \frac{1}{H(z)}
\left[ R_s(z) - I_s(z) - I_x(z)  \right],
\label{eq:dxe}
\\
(1+z)\frac{dT_k}{dz}&=2 T_k + \frac{8\sigma_T a_R T_{\gamma}^4}{3m_e
c H(z)}\frac{x_e(T_k -T_{\gamma})}{(1+f_{\rm He}+x_e)} 
 -\frac{2}{3
k_B H(z)} \frac{K_h(z)}{(1+f_{\rm He}+x_e)} .
\label{eq:dT}
\end{align}
Eq.~\eqref{eq:dxe} gives the evolution of the ionization fraction, where $R_s$ and $I_s$ are the standard primordial hydrogen recombination
rate and ionization rate respectively~\cite{1968ApJ...153....1P,1968ZhETF..55..278Z}. 
Eq.~\eqref{eq:dT} provides the evolution of the baryon temperature.
In these equations,
the contributions of the WIMP annihilation are represented
in the terms $I_x$ and $K_h$,
\begin{eqnarray}
I_x=\frac{\chi_x} {{n_{\rm H}(z)}}\frac{d^2E}{dVdt} (z),
\label{eq:Ix}
\\
K_h=\frac{\chi_h} {{n_{\rm H}(z)} E_i}\frac{d^2E}{dVdt} (z).
\label{eq:Kh}
\end{eqnarray}
where ${d^2E(z)}/{dVdt}$ is the energy injection rate of the WIMP annihilation given in Eq~\eqref{eq:dEdVdt},
$E_i$ is the ionization energy of hydrogen, $\chi_x$ and $\chi_h$ are the fractions of energy used for the ionization and heating of the cosmic plasma, respectively. In Eqs.~\eqref{eq:Ix} and \eqref{eq:Kh}, 
for simplicity, we take the on-the-spot approximation
in which the injected energy is assumed to be absorbed into the cosmic plasma and, instantaneously, consumed to ionize and heat the plasma \cite{2005PhRvD..72b3508P}.
The functions $\chi_x$ and $\chi_h$ mainly depend on the ionization fraction of the Universe, $x_e$, and we use the fitting formulae given in Ref.~\cite{2017JCAP...03..043P} which is based on the result in Ref.~\cite{Galli:2013dna}.

\begin{figure}
 \begin{tabular}{cc}
 \begin{minipage}{0.45\hsize}
  \begin{center}
    \includegraphics[width=75mm]{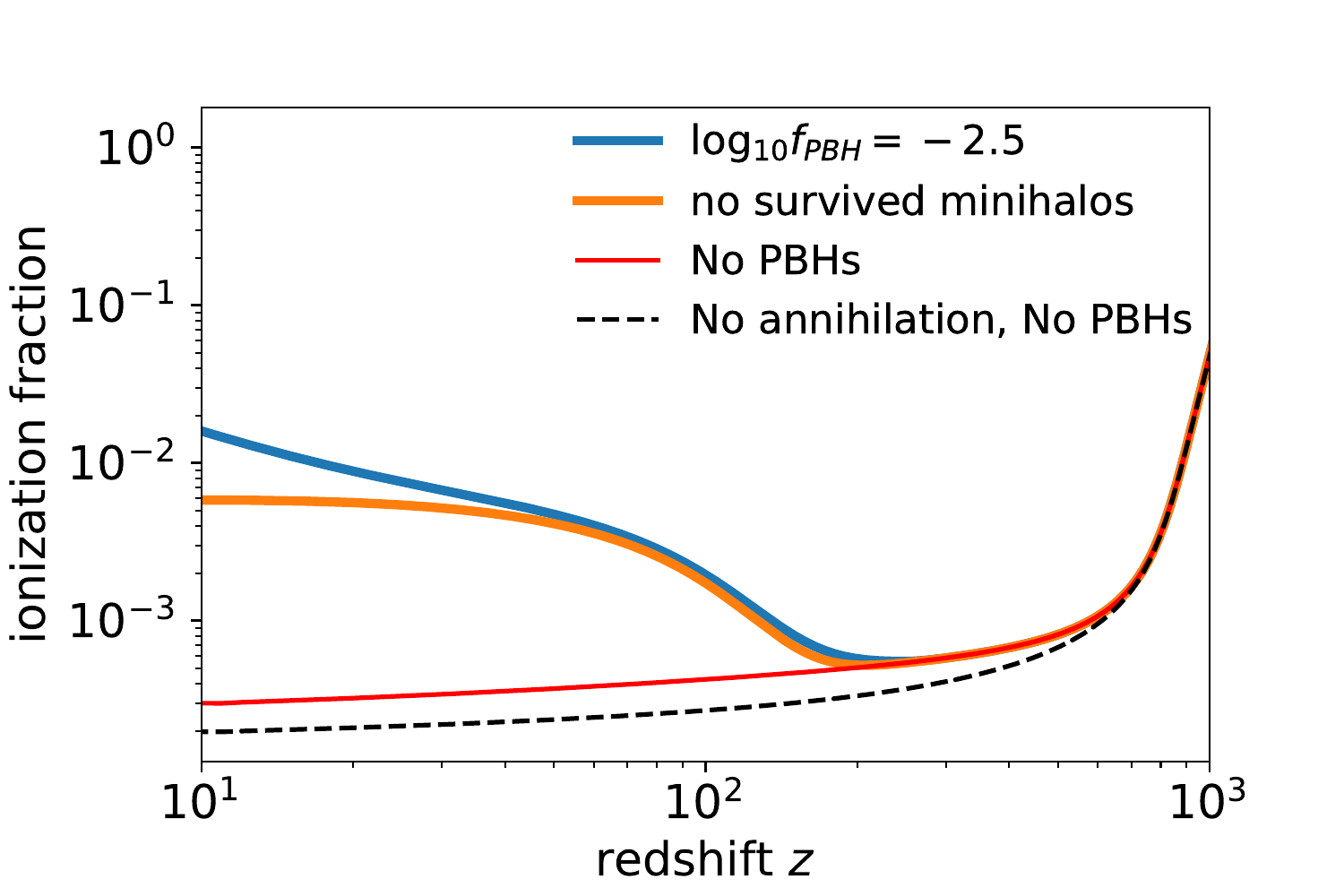}
  \end{center}
 \end{minipage}
 \begin{minipage}{0.45\hsize}
  \begin{center}
    \includegraphics[width=75mm]{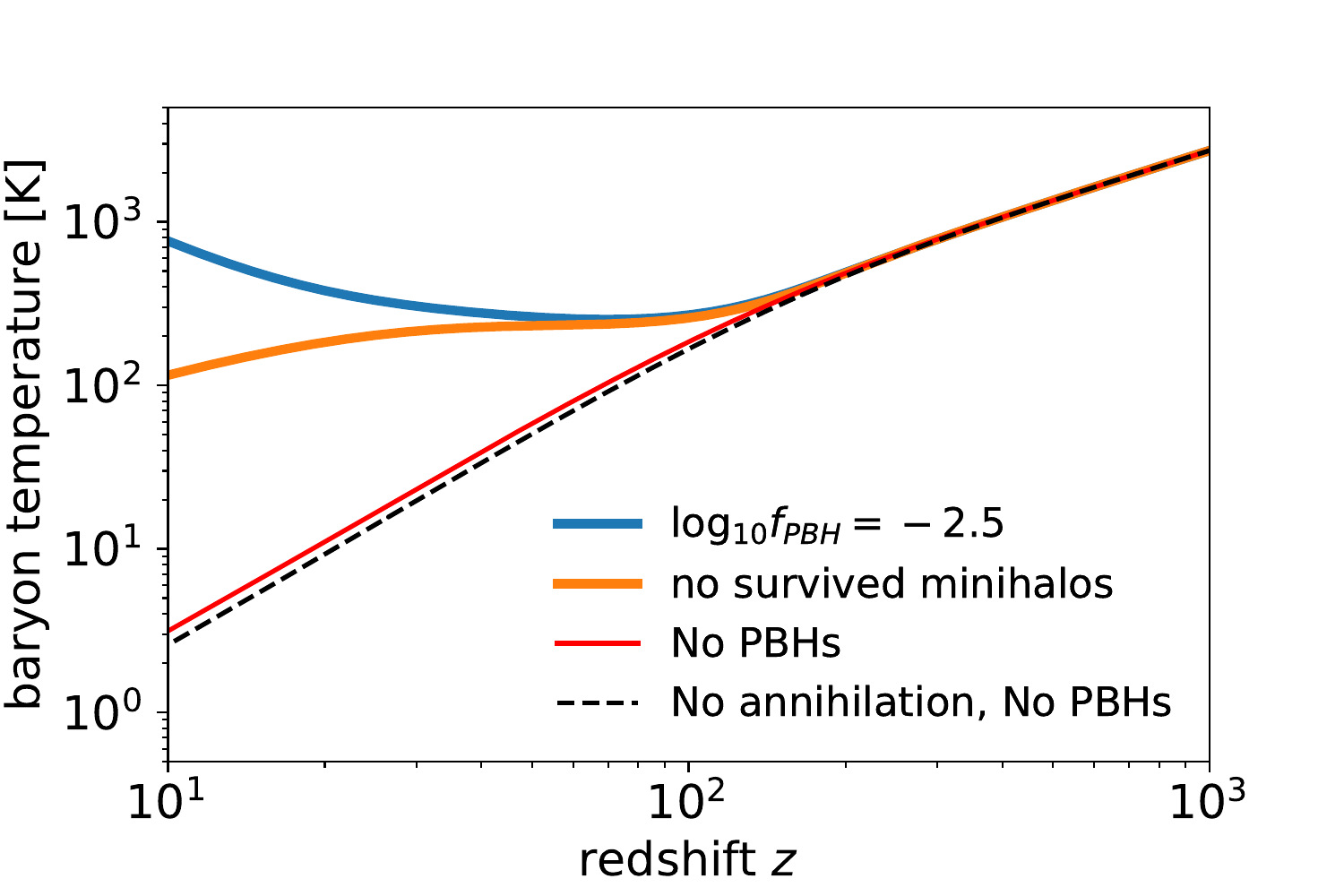}
  \end{center}
 \end{minipage}
 \end{tabular}
   \caption{The ionization fraction~(left panel) and the baryon temperature~(right panel)
   as functions of a redshift. We set $ f_{\rm PBH} = 10^{-2.5}$. The blue line represents the total contribution from DM, while the orange line is for the case without survived minihalos~($f_{\rm coll, mh} =0$). For reference, we plot the evolution without PBHs and without annihilation.
} 
  \label{fig:xe_temp}
\end{figure}

\begin{figure}
 \begin{tabular}{cc}
 \begin{minipage}{0.45\hsize}
  \begin{center}
   \includegraphics[width=75mm]{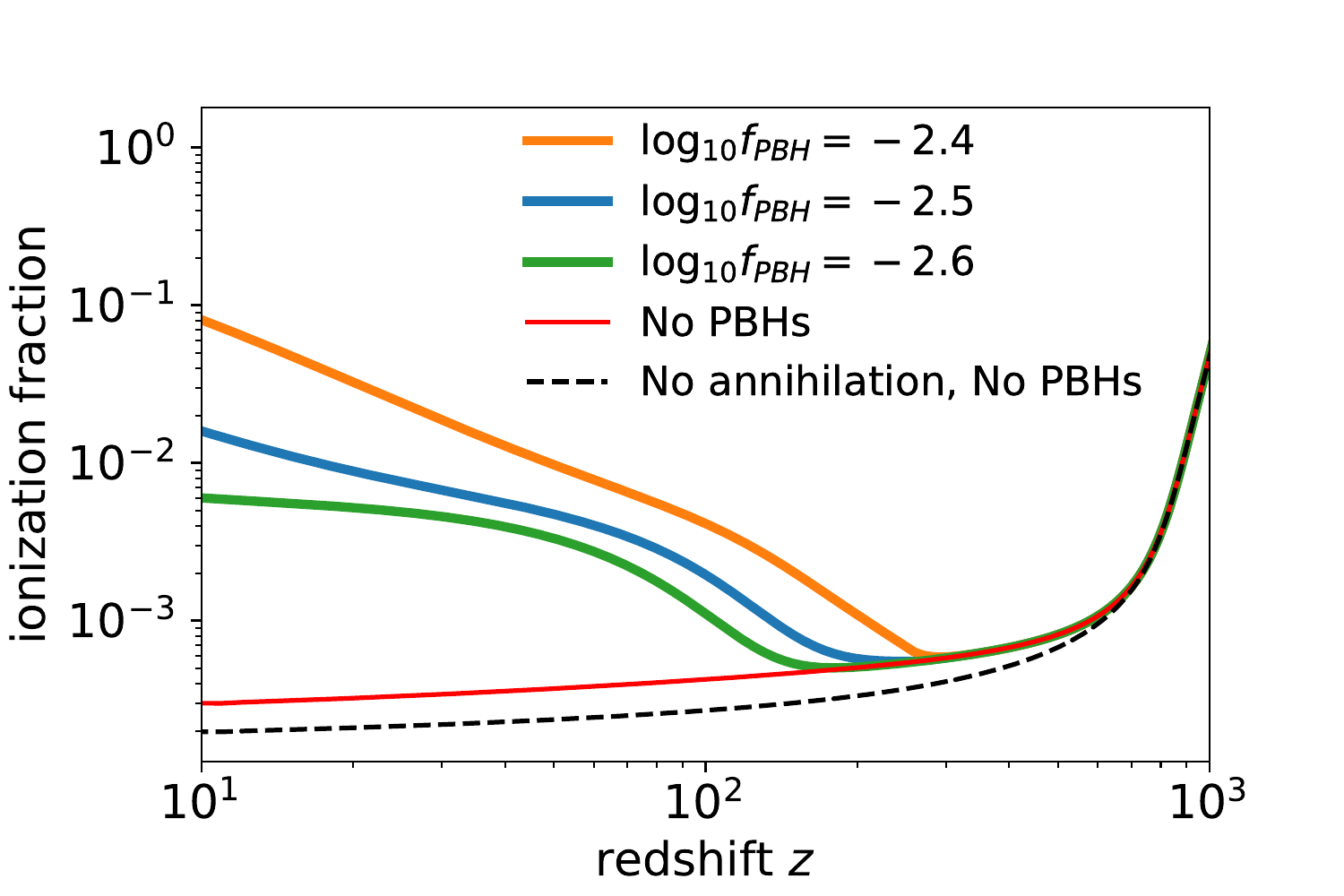}
  \end{center}
 \end{minipage}
 \begin{minipage}{0.45\hsize}
  \begin{center}
   \includegraphics[width=75mm]{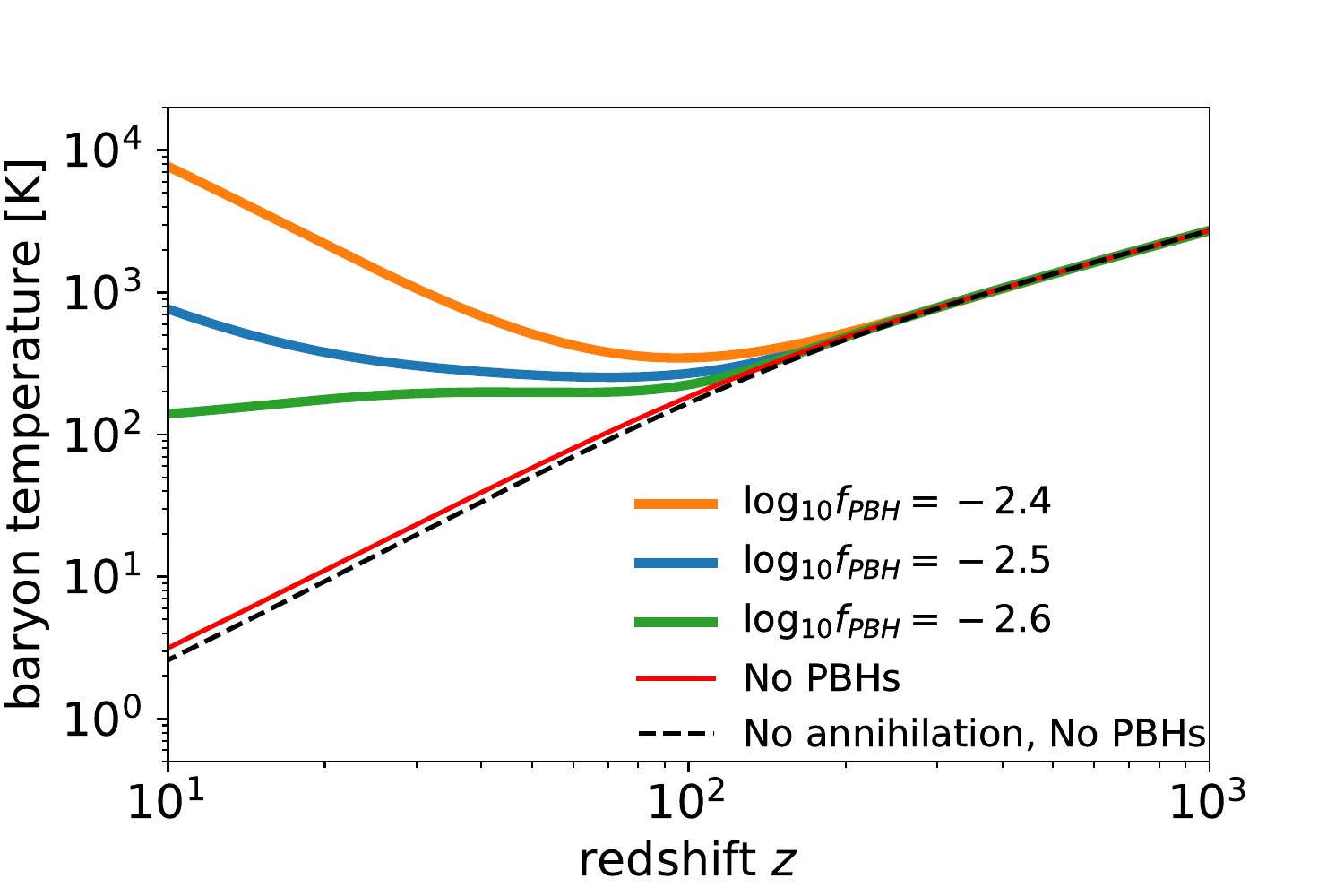}
  \end{center}
 \end{minipage}
 \end{tabular}
   \caption{The dependence on the PBH abundance $f_{\rm PBH}$ in the ionization fraction~(left panel) and the baryon temperature~(right panel). From the top to the bottom, the solid lines represent the cases with 
   $ f_{\rm PBH} = 10^{-2.4}$, $ f_{\rm PBH} = 10^{-2.5}$ and $ f_{\rm PBH} = 10^{-2.6}$.}
  \label{fig:xe_temp_fpbh}
\end{figure}

The CMB bounds depend on the combination of parameters $f_{\rm ann}\langle \sigma v \rangle /m_{\chi} $ rather than separately on each of the DM mass and annihilation cross section, and the effects of DM annihilation are conventionally parameterized by a quantity $f_{\rm ann}\langle \sigma v \rangle /m_{\chi}$. The current CMB bound is of order $f_{\rm ann}\langle \sigma v \rangle / m_{\chi}\lesssim 4 \times 10^{-28} \rm cm^3 /s /GeV$, and the canonical thermal WIMP relic value $\langle \sigma v \rangle = 3 \times 10^{-26} \rm cm^3/s$ for the DM with s-wave annihilation, for instance, can be already excluded for $m_{\chi}\lesssim 10$ GeV for a typical range of $f_{\rm ann}\sim {\cal O}(0.1)$ \cite{Slatyer:2015jla}. We for the illustration purpose use $f_{\rm ann}\langle \sigma v \rangle /m_{\chi} = 3 \times 10^{-28} \rm cm^3 /s /GeV$ as our fiducial value in the figures unless stated otherwise.\footnote{Using for instance the value $f_{\rm ann}\langle \sigma v \rangle /m_{\chi} = 3 \times 10^{-29} \rm cm^3 /s /GeV$, instead of $3 \times 10^{-28} \rm cm^3 /s /GeV$, does not affect our following discussions, even though the constraint on $f_{PBH}$ becomes less tight by a factor $\sim$ 5 as mentioned in the next section.} 
We solve Eqs.~\eqref{eq:dxe} and \eqref{eq:dT}
by using the public code {\tt HyRec}~\cite{2011PhRvD..83d3513A}
in which the primordial helium contribution is also included.
In Fig.~\ref{fig:xe_temp}, we show the effect of the WIMP annihilation in the mixed DM scenario
on the ionization fraction (left panel) and the baryon temperature~(right panel).
In this figure using $f_{\rm ann}\langle \sigma v \rangle /m_{\chi} = 3 \times 10^{-28} \rm  cm^3 /s/ GeV$, the blue line represents the evolution with $\log_{10} f_{\rm PBH} = -2.5$, and, for comparison, we also plot the results 
for no PBH case~(a conventional adiabatic perturbation scenario without PBH isocurvature perturbations) and for no DM annihilation case with the thin red line and black dashed line, respectively. In order to illustrate the impact of minihalo contributions, we also plot the evolution for the case with no survived minihalos~(i.e., $f_{\rm coll,mh}=0$)
in the orange line in Fig.~\ref{fig:xe_temp}.
The figure tells us that the survived minihalo contribution is significant in the late universe. 
When $\log_{10} f_{\rm PBH} = -2.5$,
the mass dispersion is small
$\sigma (M_{\rm min}, z_*) =0.4~(<\delta_c\sim 1.686)$ at $z_* = 250$ and the collapsed matter fraction locked in the survived minihalos would be consequently small.
As a result, the survived minihalo contribution is not significant until a later epoch $z\sim 50$. 
As the universe expands further, the core density in survived minihalos relatively becomes significantly large, compared with the background density or the DM halo density.
Around $z\sim 50$, the boost factor of survived minihalos cannot be negligible and finally dominates.
The ionization and heating from the annihilation in survived minihalos can significantly affect the Thomson optical depth of the CMB and 21-cm signals in the dark ages. We also show the evolution of the ionization fraction and baryon temperature for different $f_{\rm PBH}$ values in Fig. \ref{fig:xe_temp_fpbh}. A large $f_{\rm PBH}$ induces the early structure formation and the effect of the halo contribution start to become important at higher redshifts. The survived minihalo contribution becomes smaller when $f_{\rm PBH}$ becomes smaller, and we found that, when $
f_{\rm PBH} \lesssim 10^{-2.6}$, 
the contribution from the survived minihalos, the first term in the right-hand side of Eq.~\eqref{eq:dEdVdt_halo},
is negligible in the thermal evolution of the Universe.

\section{cosmological constraints on the PBH abundance}
\label{sec5}
\subsection{CMB anisotropy}

As shown in the last section,
the WIMP annihilation can be enhanced in the mixed DM scenario
and hence induce the early reionization.
One of cosmological probes on such early reionization is 
the CMB anisotropy measurement.
In particular, the CMB polarization is created due to the scattering of free electrons
in the propagation of CMB photons from the last scattering surface. The free electron density indeed increases for a bigger $f_{\rm PBH}$ as shown in  Fig.~\ref{fig:xe_temp_fpbh}.

\begin{figure}
  \begin{center}
  \includegraphics[width=10.0cm]{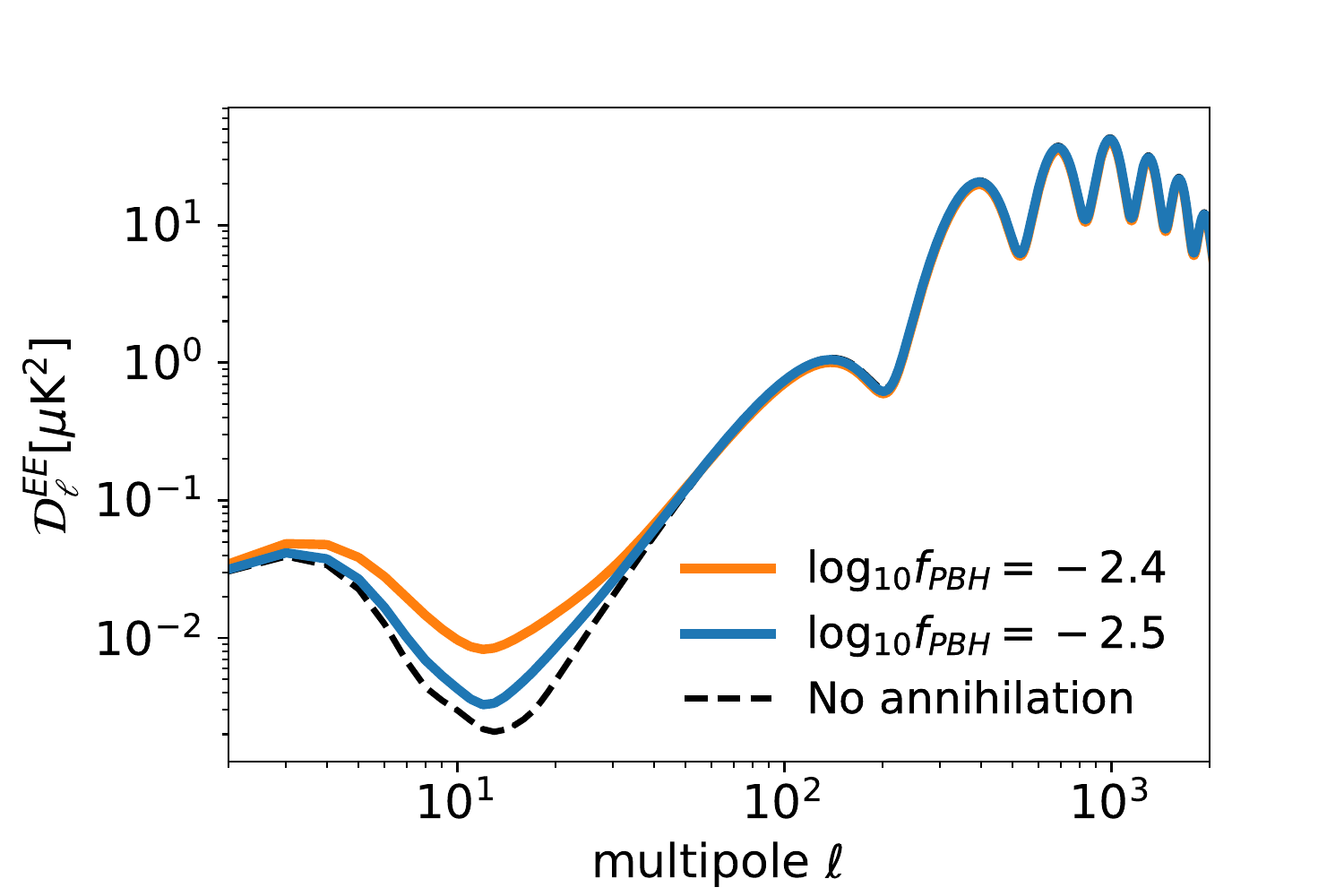}
  \caption{Angular power spectrum of CMB E-mode polarization in the mixed DM scenario of PBHs and self-annihilating DM. From the top to the bottom, the solid lines are the power spectra for 
   $ f_{\rm PBH} = 10^{-2.4}$ and $ f_{\rm PBH} = 10^{-2.5}$. For reference, we show the angular power spectrum without the DM annihilation in the dashed line.}
  \label{fig:clee}
\end{center}
\end{figure}

Taking into account the enhancement of the ionization fraction, we evaluate the CMB polarization anisotropy by the public Boltzmann code~{\tt CLASS}~\cite{2011JCAP...07..034B}.
Fig.~\ref{fig:clee} shows
the angular power spectra of the CMB E-mode polarization
for different $f_{\rm PBH}$ values. We adopt the "tanh"-shape reionization history with $z_{\rm reio} = 7.68$ \cite{ag2018} to include the standard reionization scenario driven by first stars and galaxies.
The figure tells us that the early reionization due to a large $f_{\rm PBH}$ increases the CMB polarization signal represented by the change in the reionization bump. 
The detailed CMB anisotropy measurement hence can 
provide a constraint on the PBH abundance. In order to obtain the constraint from the current CMB measurement,
we perform the MCMC analysis using {\tt Monte Python}~\cite{Audren:2012wb}. In this analysis we use the baseline likelihood~(TTTEEE-lowl-lowE) from the Planck 2018 data release~\cite{Aghanim:2019ame} and adopt the standard six cosmological parameters
with the PBH abundance $\{\omega_b, \omega_d, 100\theta_s, \ln (10^{10}A_s), n_s, z_{\rm reio}, f_{\rm PBH}\}$.

\begin{figure}
  \begin{center}
  \includegraphics[width=10.0cm]{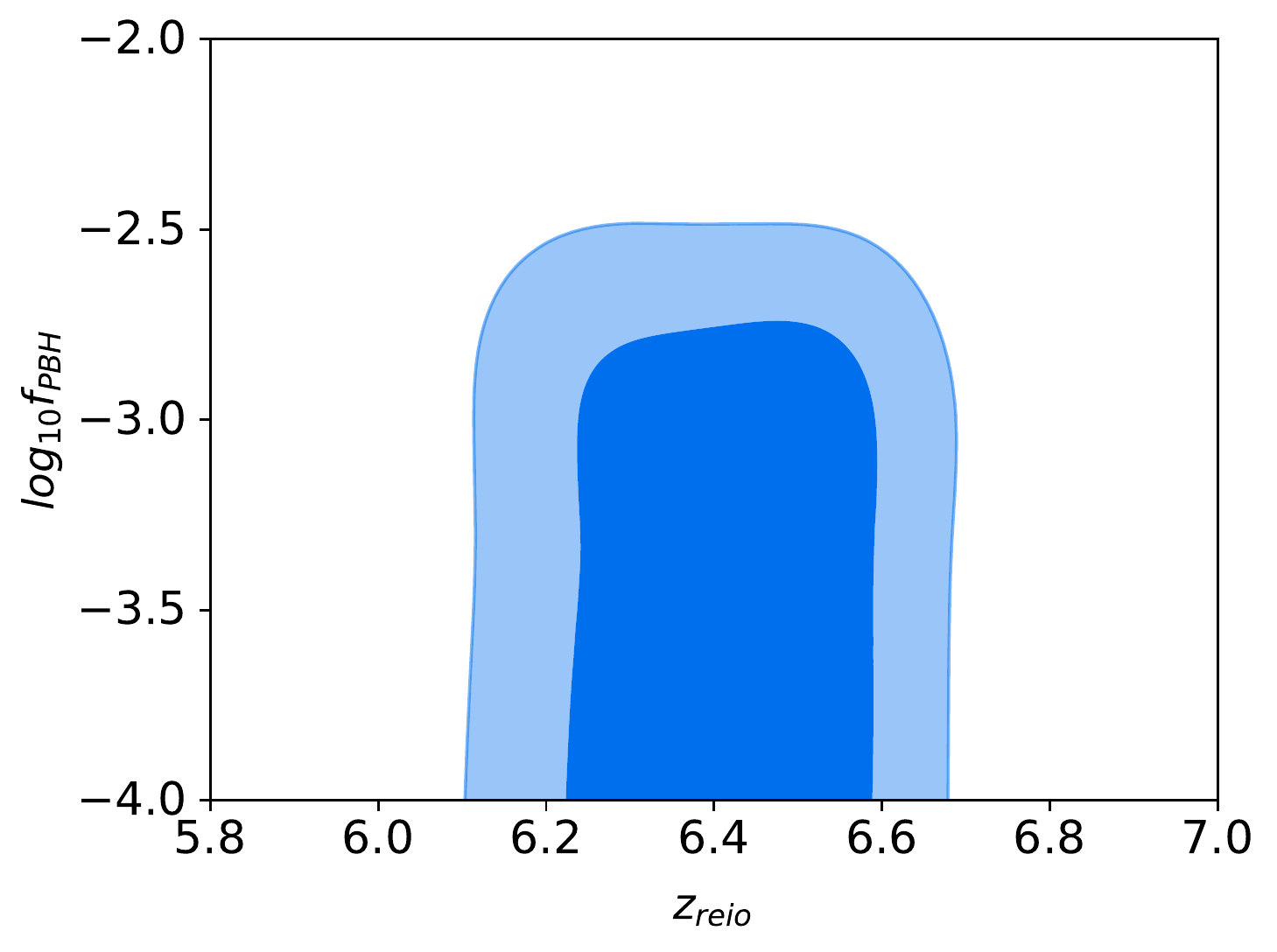}
  \caption{Two-dimensional contour plot for $z_{\rm reio}$ and $f_{\rm PBH}$ from the MCMC analysis with Planck 2018 data. Here we set $f_{\rm ann}\langle \sigma v \rangle /m_{\chi} = 3 \times 10^{-28} \rm  cm^3 /s/ GeV$. }
  \label{fig:zre_fpbh}
\end{center}
\end{figure}

Fig~\ref{fig:zre_fpbh} shows 2-D contour plot for $z_{\rm reio}$ and $f_{\rm PBH}$. 
The annihilation enhancement occurs when the fluctuations grow enough to produce halos and induces the abrupt early reionization.
PBH isocurvature fluctuations lead to the early structure formation and significantly enhance such effects.
The MCMC analysis is sensitive to the 
peak height of the acoustic part in 
the temperature anisotropy, which is 
proportional to $A_s e^{-2 \tau}$ where 
$\tau$ is the optical depth to the last scattering surface.
This optical depth $\tau$ is 
a sum of contributions from the epoch soon after the last scattering surface~($200 \lesssim  z \lesssim 1100$), the early reionization epoch~($20 \lesssim  z \lesssim 200$) 
and the late reionization epoch~($z\lesssim20$) (this late reionization is parameterized by "tanh"-shape reionization model centered at $z=z_{\rm reio}$).
The DM annihilation in halo structures largely affect
the ionization fraction in the early reionization part by the abrupt increase of the reionization fraction
even if $f_{\rm PBH} =0 $ (because of the conventional adiabatic perturbations).
Not to affect the total $\tau$, a smaller $z_{\rm reio}$ is preferable.
We also note the best fit value of $A_s$ does not change dramatically in the presence of PBH in our analysis partly because the uncertainty in the polarization data is larger than that of the temperature anisotropy. Hence the MCMC tends to vary $z_{\rm reio}$ rather than $A_s$ to keep the amplitude of  $A_s e^{-2 \tau}$.
In other words, to maintain the peak height in the CMB temperature anisotropy,
the smaller contribution of the conventional tanh reionization model is preferable to compensate the additional contribution from WIMP annihilation reionization. This hence leads to the smaller $z_{\rm reio}$ compared with $z_{\rm reio}
=  7.67 \pm 0.73$ in the standard $\Lambda$CDM case~\cite{ag2018}~(without including DM annihilation).

Our MCMC analysis provides the constraint on the PBH abundance,
$ f_{\rm PBH} <10^{ -2.66}$ for $f_{\rm ann}\langle \sigma v \rangle /m_{\chi} = 3 \times 10^{-28} \rm  cm^3 /s/ GeV$,
at the 95\%~confidence level. 
For the DM annihilating into the SM particle pairs, a typical range of $f_{\rm  ann}$ is of order $f_{\rm ann} \approx 0.2 - 0.6$ \cite{Slatyer:2015jla} and we can obtain the constraint
\begin{align}
 f_{\rm PBH} <10^{ -2.66} \quad  {\rm for~}\langle \sigma v \rangle /m_{\chi} \sim 10^{-27} \rm  cm^3 /s/ GeV,
\label{eq266}
\end{align}
at the 95\%~confidence level. As mentioned in the last section,
the survived minihalo contribution is negligible when $\log_{10} f_{\rm PBH} < -2.6$. Therefore, this constraint  given by Eq. (\ref{eq266}) does not contain the model uncertainty related to the abundance
of survived minihalos.
On the other hand, the constraint strongly depends on the annihilation rate $f_{\rm ann} \langle \sigma v \rangle /m_{\chi}$. 
When we take $f_{\rm ann}\langle \sigma v \rangle /m_{\chi} =  3 \times 10^{-29} \rm  cm^3 /s/ GeV$,
the constraint is relaxed to 
$f_{\rm PBH} < 10^{-2.38}$ at the 95\%~confidence level.

\begin{figure}
  \begin{center}
  \includegraphics[width=10.0cm]{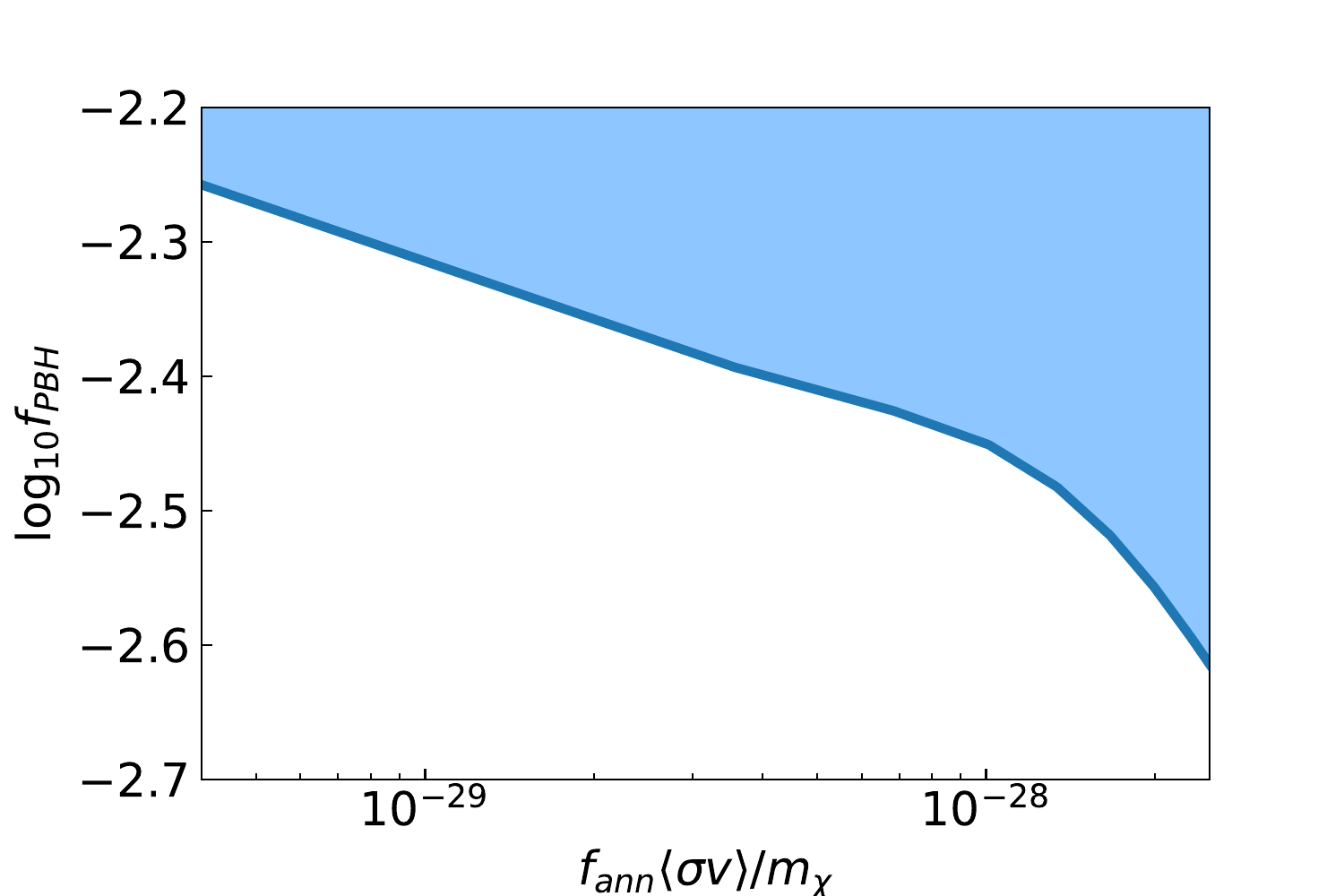}
  \caption{The constraints on $f_{\rm PBH}$
  for different values of $f_{\rm ann} \langle \sigma v \rangle /m_{\chi}$.
  The colored region is excluded by the CMB anisotropy measurement.
}
  \label{fig:const_fpbh}
\end{center}
\end{figure}

Fig.~\ref{fig:const_fpbh} is the summary of our constraint on $f_{\rm PBH}$.
The colored region represents the excluded parameter region for each $f_{\rm ann} \langle \sigma v \rangle /m_{\chi}$
by our analysis.
As mentioned in the previous section, when $ f_{\rm PBH} \gtrsim 10^{-2.6}$, 
the contribution from the survived minihalos, the first term in the right-hand side of Eq.~\eqref{eq:dEdVdt_halo}, becomes significant.
The minihalo contribution becomes too big for $f_{\rm ann} \langle \sigma v \rangle /m_{\chi}\gtrsim 10^{-28} \rm cm^3/s/GeV$ if $ f_{\rm PBH} \gtrsim 10^{-2.6}$, and the excluded region for $f_{\rm ann} \langle \sigma v \rangle /m_{\chi}\gtrsim 10^{-28} \rm cm^3/s/GeV$ represents the parameter sets where the survived minihalo contribution cannot be ignored compared with the smooth background contribution. For $f_{\rm ann} \langle \sigma v \rangle /m_{\chi}\lesssim 10^{-28} \rm cm^3/s/GeV$, there is room for the minihalo contribution to dominate the smooth background contribution, and the slope of the constraint becomes smaller compared with that for a higher $f_{\rm ann} \langle \sigma v \rangle /m_{\chi}$ for which there is little room for the minihalo contribution.

\subsection{CMB $y$-type distortion}

\begin{figure}
  \begin{center}
  \includegraphics[width=10.0cm]{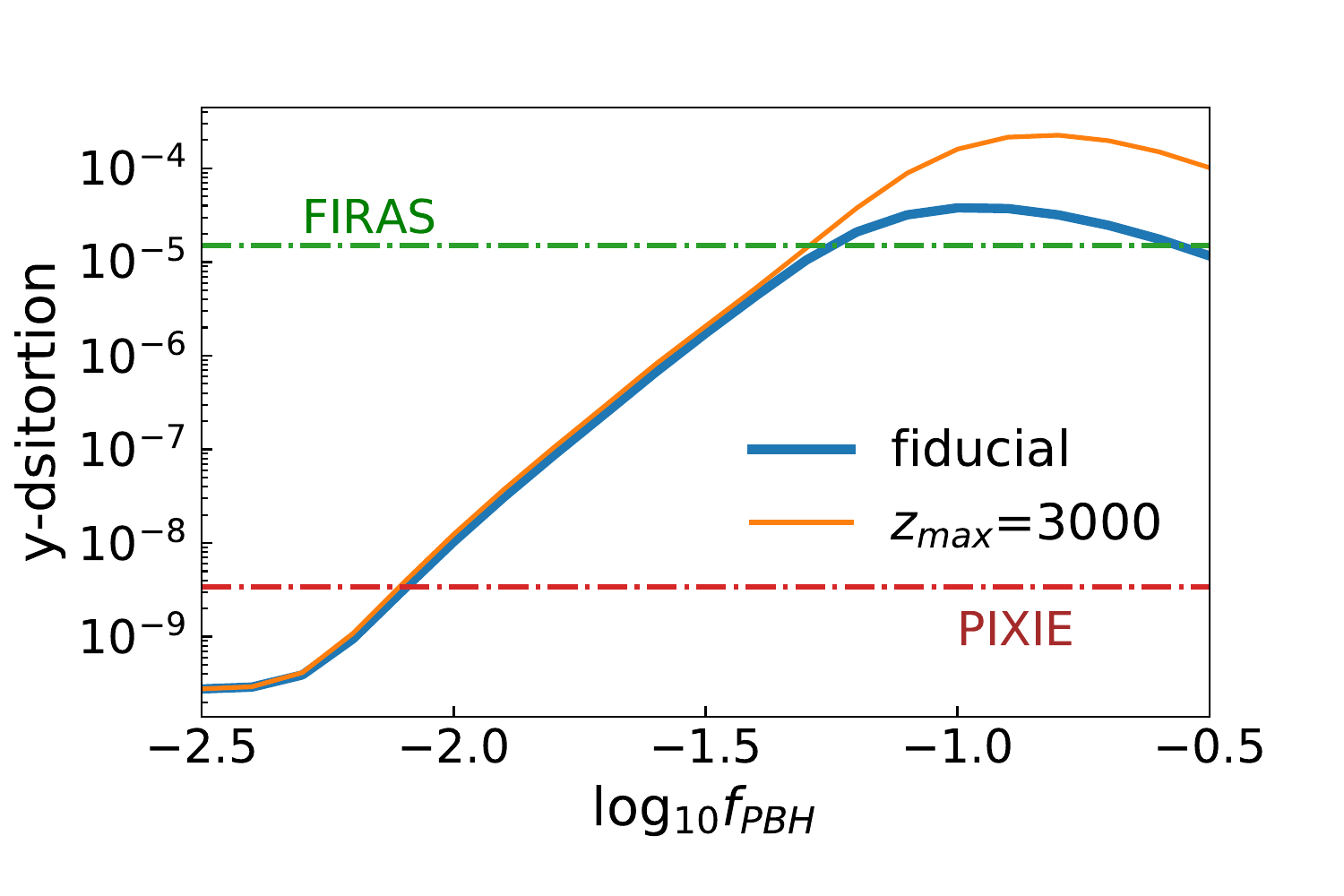}
  \caption{CMB $y$-type distortion as a function of $f_{\rm PBH}$. Our fiducial case is represented in a thick blue line with $f_{\rm ann}\langle \sigma v\rangle/m_{\chi}=3\times 10^{-28}~\rm cm^3/s/GeV$ and $z_{\rm max}=2000$. To represent the dependence on $z_{\rm max}$,
  we plot the result with $z_{\rm max} = 3000$ in a thin orange line.}
  \label{fig:cmb_dis}
\end{center}
\end{figure}

The CMB distortion can offer another powerful probe on the thermal history of the Universe~\cite{2014PTEP.2014fB107T,2014MNRAS.438.2065C,2014arXiv1405.6938C}.
The injected energy from the DM annihilation creates the deviation from the black-body spectrum in the CMB energy spectrum.
In particular, the enhancement of the structure formation due to the PBHs can result in the enhanced $y$-type distortion.
The CMB $y$-type distortion can be estimated by~\cite{2016MNRAS.460..227C}
\begin{equation}
y \approx \frac{1}{4} \int^{z_{\mu y}} _{z_0} \frac{dz}{(1+z)} \frac{\dot Q^{\rm heat}}{H(z) \rho_{\gamma}(z)}   ,
\end{equation}
where $\rho_{\gamma} (z)$ is the CMB energy density at $z$
and we set $z_{\mu y} = 5\times 10^4$.
Before $z_{\mu y}$, $\mu$-type distortion is created instead of $y$-type distortion.
We take $z_0 = 200$ because the energy transfer from baryons to CMB becomes inefficient after $z_0$ \cite{2016MNRAS.460..227C}.
Fig. ~\ref{fig:cmb_dis} shows the created $y$-type distortion as a function of the PBH fraction, $f_{\rm PBH}$. The blue line represents our fiducial case with $f_{\rm ann}\langle \sigma v \rangle /m_{\chi}=3\times 10^{-28} \rm cm^3 /s/ GeV$.
As $f_{\rm PBH}$ increases, the larger PBH isocurvature fluctuations can lead to the larger $y$-type distortion. However, when $f_{\rm PBH} >0.1$, $y$-type distortion decreases with an increasing $f_{\rm PBH}$ because the amount of annihilating DM is not sufficient to generate the large CMB distortion (we remind readers that $\Omega_{\rm DM} = \Omega_{\rm PBH } + \Omega_{\chi}$).

One of the model parameters, which strongly affects $y$-type distortion, is
$z_{\rm max}$ which represents the maximum redshift for halo formations. Although we conservatively set $z_{\rm max} = 2000$ in the fiducial model,
halos can form when the density fluctuations grow sufficiently after the epoch of the matter-radiation equality. 
In order to evaluate the impact of $z_{\rm max}$ on $y$-type distortion, we calculate $y$-type distortion with $z_{\rm max}=3000$
and plot the result in the thin orange line in Fig.~\ref{fig:cmb_dis}.
When $f_{\rm PBH}$ is large, $y$-type distortion strongly depends on $z_{\rm max}$. 
A larger $f_{\rm PBH}$ induces an earlier formation of minihalos whose DM density $(\propto (1+z_{\rm f})^3)$ can be big, and a larger $y$-type distortion can be created from the DM annihilation with a higher $z_{\rm max}$.
However, when $\log_{10} f_{\rm PBH} < -1.5$, the halo formation at $z>2000$ is not so efficient and the effect on $y$-type distortion is small even if $z_{\rm max } =2000$ is changed to $z_{\rm max } =3000$ in our analysis.
The current limit of $y$-type distortion is given by COBE/FIRAS,~$y < 1.5 \times 10^{-5}$~\cite{Fixsen:1996nj}, while the projected 1-$\sigma$ detection sensitivity for the next-generation PIXIE-like experiment \cite{2011JCAP...07..025K} is $\sim 3.4 \times 10^{-9}$~\cite{2019arXiv190901593C}.
As discussed for the CMB anisotropy, $f_{\rm PBH}$ is constrained as $f_{\rm PBH} <10^{ -2.66}$ 
from Planck 2018 data. 
The PBH abundance which can create the PIXIE detection level of $y$-type distortion is already ruled out by Planck observation.
This is partly because the spectral distortion is affected by the DM annihilation energy going into the abundant CMB photons while the CMB anisotropy bounds mainly come from the ionization fraction which concerns the DM annihilation energy going into the gas. It is hence reasonable to expect the latter can give the tighter bounds on $f_{\rm PBH}$ considering the much bigger number density of photons than baryons, in addition to the current poor precision on the CMB spectral distortion measurements compared with the CMB anisotropy data.

\subsection{Global 21-cm signal}
\label{sec6}
Redshifted 21-cm signals can probe the IGM evolution of the thermal and ionization history. The multifrequency radio observation can measure the 21-cm signals from different redshifts corresponding to different observation frequencies.
The measured quantity in 21-cm observations is the 
so-called differential brightness temperature, which is the deviation of the redshifted 21-cm brightness temperature from the CMB temperature.
Similarly to CMB, both global~(sky-averaged) signal and spatial fluctuations of differential brightness temperature can provide valuable information about our Universe~(for a comprehensive review, see Ref.~\cite{2006PhR...433..181F}). Here we focus on the global signal.

\begin{figure}
 \begin{tabular}{cc}
 \begin{minipage}{0.45\hsize}
  \begin{center}
   \includegraphics[width=75mm]{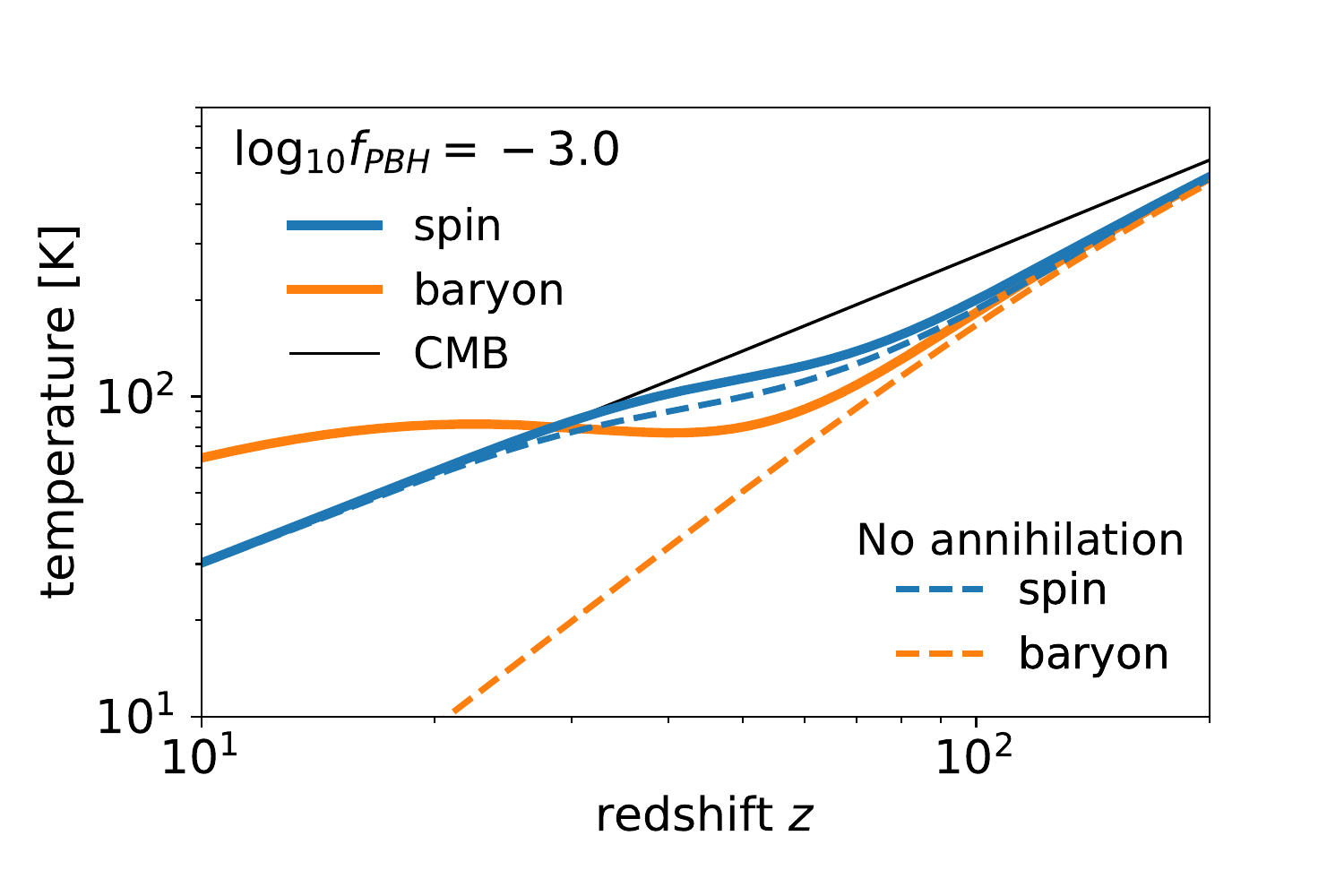}
  \end{center}
 \end{minipage}
 \begin{minipage}{0.45\hsize}
  \begin{center}
   \includegraphics[width=75mm]{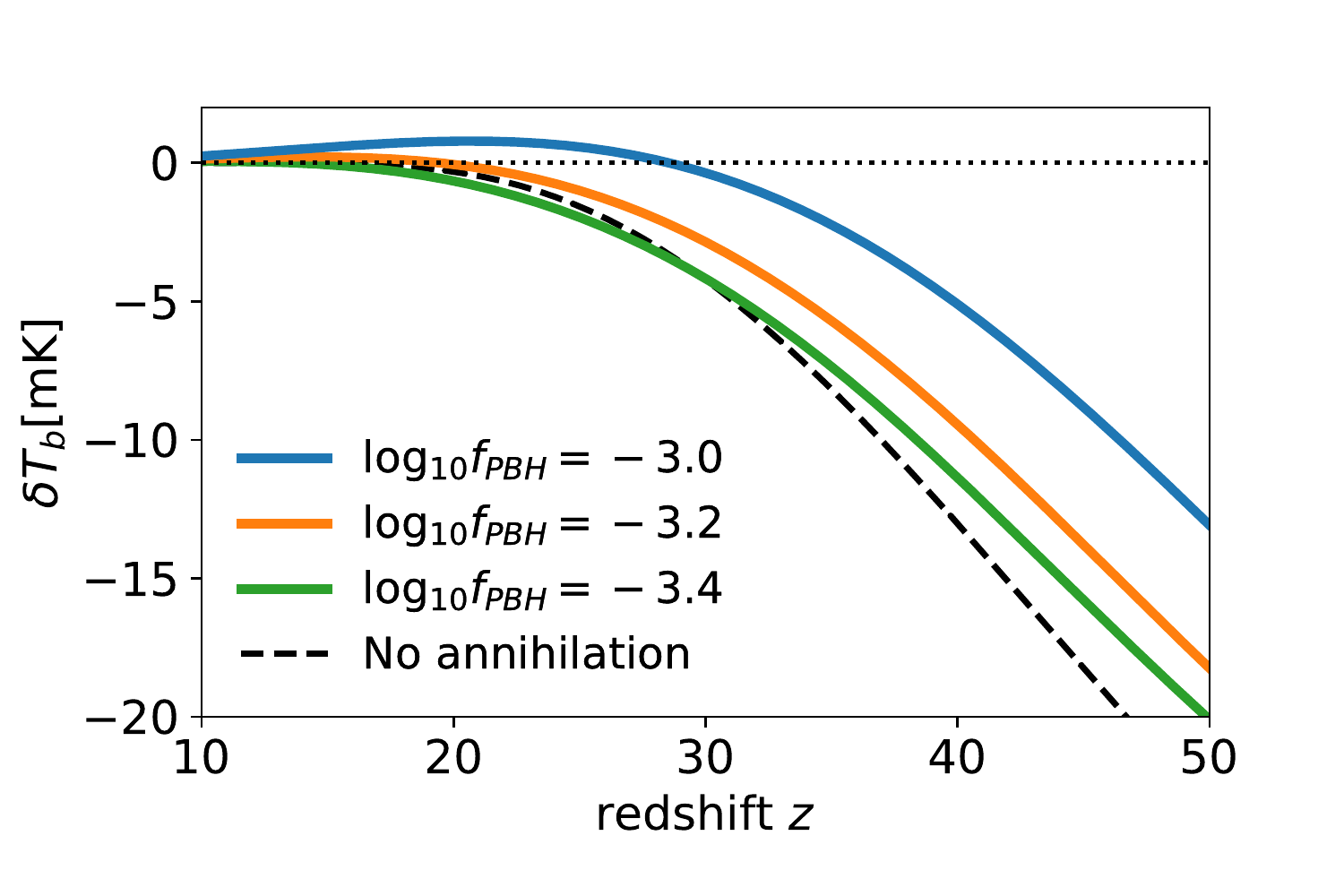}
  \end{center}
 \end{minipage}
 \end{tabular}
   \caption{{\it Left panel}: The spin and baryon temperatures as functions of redshifts. The solid blue and orange lines represent the spin and baryon temperatures for $ f_{\rm PBH} = 10^{-3.0}$. For comparison, we plot their evolutions in the no annihilation DM case in the dashed lines. 
   {\it Right panel}: 
   The evolution of the global differential brightness temperature with different $f_{\rm PBH}$. 
   From top to bottom, the solid lines are for $ f_{\rm PBH} = 10^{-3.0}$, $f_{\rm PBH} =10^{ -3.2}$ and $ f_{\rm PBH} =10^{ -3.4}$. For reference, the dashed line shows the evolution in the case of no annihilation DM.
}
  \label{fig:dTb}
\end{figure}

The global differential brightness temperature from a redshift~$z$ is provided by~\cite{1997ApJ...475..429M,2010PhRvD..82b3006P}
\be
\label{eq:differential_temp}
\delta T_b (z)=
\frac{3}{32 \pi} \frac{hc^3 A_{10}}{k_B \nu_0^2} \frac{x_{\rm HI} n_H}{(1+z)^2 (dv_{||}/dr_{||})} \left( 1-\frac{T_\gamma}{T_S}\right),
\ee
where $dv_{||}/dr_{||}$ is the gradient of the proper velocity along the line of sight, which is the Hubble expansion contribution in the global signal. $T_S$ is the spin temperature of the neutral hydrogen hyperfine structure.
The spin temperature is determined by the balance in the processes of the hyperfine excitation and de-excitation,
\be
T_S = \frac{T_\gamma + y_{\rm kin} T_k}{1+y_{\rm kin}},
\ee
where $y_{\rm kin}$ represents the efficiency ratio between
the absorption of CMB photons and the thermal collisions
in the hyperfine transition. We adopt the approximated analytical form of $y_{\rm kin}$ in~Ref.~\citep{2006ApJ...637L...1K}.
In Eq.~\eqref{eq:differential_temp}, we do not include the contribution from Ly-$\alpha$ coupling,
because it is not important until the beginning of the first star formation.
In the left panel of Fig.~\ref{fig:dTb}, we plot the evolution of spin and baryon temperatures with 
$ f_{\rm PBH} = 10^{-3}$ in solid lines.
For comparison, we also plot the evolutions in the standard cosmology case~("no annihilation" case). Here we do not consider other heating sources including stars and galaxies.
For $ f_{\rm PBH} = 10^{-3}$,
the annihilation from DM halos becomes effective below $z \sim 100$ when baryons are heated up and the thermal evolution of baryons deviate from the adiabatic evolution, $T_k \propto (1+z)^2$, in "no annihilation" case. 
Due to heated baryons, the spin temperature also becomes large with $ f_{\rm PBH} = 10^{-3}$, compared with "no annihilation" case.
The spin temperature lies between the CMB and baryon temperatures. 
Therefore, after the baryon temperature becomes larger than the CMB temperature, the spin temperature also becomes higher than the CMB temperature for $ f_{\rm PBH} = 10^{-3}$. On the other hand, the spin temperature never exceeds the CMB temperature in "no annihilation" case.

The right panel of Fig.~\ref{fig:dTb} shows the global differential brightness temperature as a function of a redshift.
The sign of the differential brightness temperature depends on the spin temperature. When the spin temperature is smaller than the CMB temperature, the differential brightness temperature is negative and observed as absorption signals on the CMB frequency spectrum.
As $f_{\rm PBH}$ decreases, the effects of the DM annihilation becomes small and the redshift, $z_{\rm tr}$, at which the global differential brightness temperature shifts from the absorption to the emission signal becomes smaller.
In the case of "no annihilation", the differential brightness temperature cannot turn to the positive side because of the lack of heating sources and the spin temperature stays lower than the CMB temperature.
Therefore, identifying $z_{\rm tr}$ observationally can provide the constraint on the heating source including the DM annihilation.

\begin{figure}
  \begin{center}
  \includegraphics[width=10.0cm]{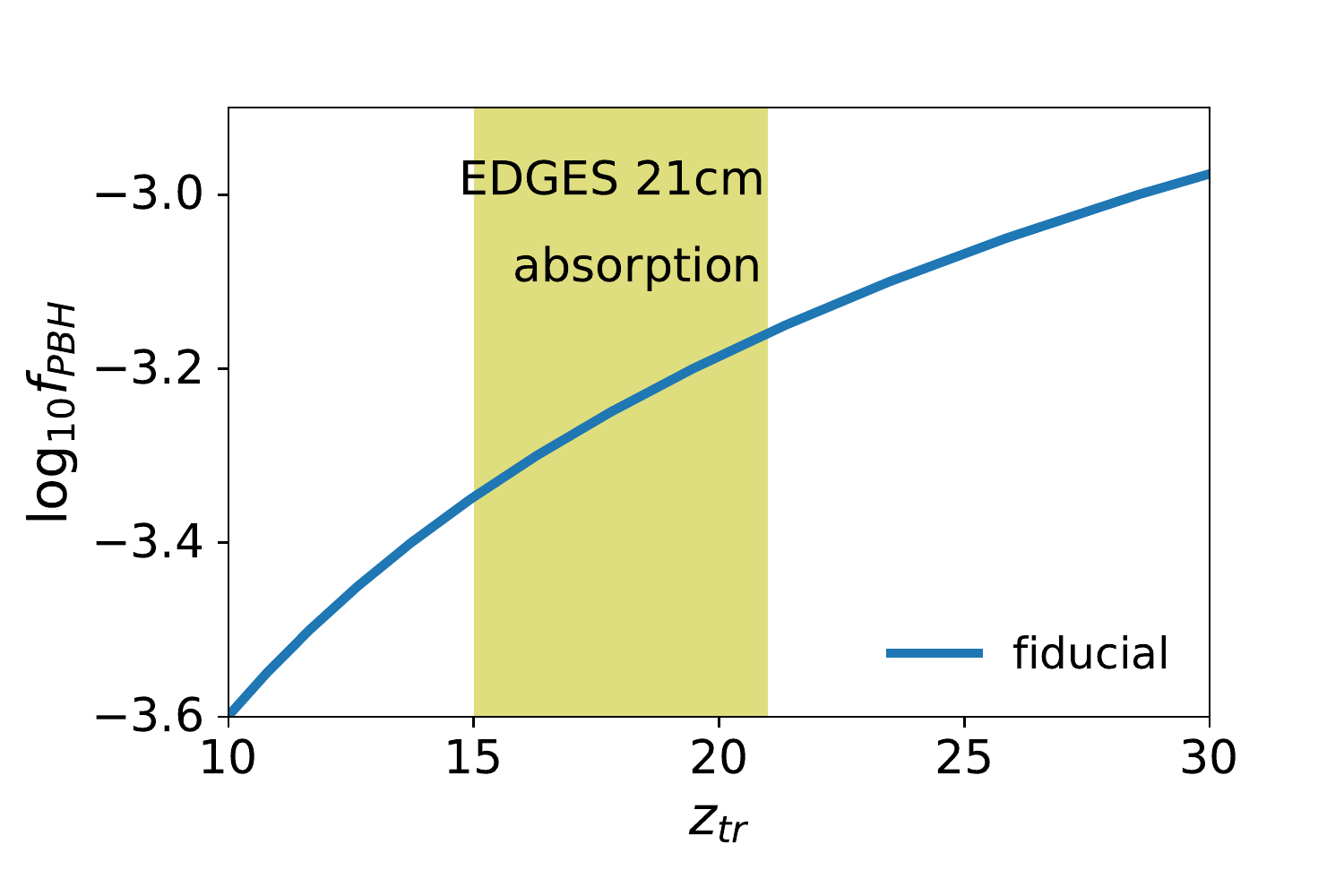}
  \caption{The dependence of the transition redshift,~$z_{\rm tr}$, from the absorption to emission on the PBH fraction $f_{\rm PBH}$.}
  \label{fig:edges-21cm}
\end{center}
\end{figure}

In Fig.~\ref{fig:edges-21cm}, we show
the relation between $z_{\rm tr}$ and $f_{\rm PBH}$. A small $f_{\rm PBH}$ provides a low $z_{\rm tr}$.
Recently the EDGES reported the detection of the global 21-cm absorption in the redshift range between $z\sim 21$ and $z\sim 15$~\cite{2018Natur.555...67B}.
This result suggests that the baryon temperature is lower than the CMB temperature until $z\sim 15$
and can limit the heating source causing the early reionization. 
As shown in Fig.~\ref{fig:edges-21cm}, $ f_{\rm PBH} > 10^{ -3.4}$ leads to the emission signal above  $z\sim 15$.
The EDGES absorption signal at $z\sim 15$ hence requires $ f_{\rm PBH} < 10^{ -3.4}$ with $f_{\rm ann}\langle \sigma v\rangle/m_\chi = 3 \times 10^{-28} {\rm cm^3/s/GeV}$.
Therefore, for $f_{\rm ann}\sim {\cal O}(0.1)$ (a typical range for the SM particle annihilation final states \cite{Slatyer:2015jla}),
our constraint is
\be
 f_{\rm PBH} < 10^{ -3.4} \quad{\rm with~} \langle \sigma v\rangle/m_\chi \sim 10^{-27} {\rm cm^3/s/GeV}.
\label{eq:fpbh_edges}
\ee
As mentioned before, when $f_{\rm PBH} < 10^{-2.6}$, the contribution from survived minihalos is negligible. Therefore, this constraint given by Eq. \eqref{eq:fpbh_edges} does not include the substructure contributions and hence does not suffer from the model parameter uncertainties related to $z_{\rm max}$ and $z_{\rm f}$.
On the other hand, $f_{\rm ann}\langle \sigma v\rangle/m_\chi$ strongly affects the constraint.
If $f_{\rm ann}\langle \sigma v\rangle/m_\chi = 3 \times 10^{-29} {\rm cm^3/s/GeV}$, the EDGES provides the constraint $f_{\rm PBH} \lesssim 10^{-2.5}$.
We also note that the baryon temperature larger than the CMB temperature can turn the absorption into the emission to result in the tight bound on $f_{\rm PBH}$, but it is not high enough to sufficiently ionize the IGM. 
  This is a reason why the bounds on $f_{\rm PBH}$ from the global 21cm signals can become tighter than those from the CMB.

\section{conclusion}

In this paper, we studied the bounds on the PBH abundance in 
the mixed DM scenarios consisting of the self-annihilating DM and PBHs. 
The Poisson distribution of PBHs can lead to the isocurvature perturbations
and consequently to the early structure formation. The enhancement of the DM annihilation occurs in
those early formed dense halos and
modifies the ionization and temperature evolutions of baryons.
Such modifications affect early Universe observables such as the CMB and 21cm signals.
In order to obtain the constraint on the PBH abundance from CMB observations, 
we performed the MCMC analysis with the latest Planck data.
The obtained constraint is  $f_{\rm PBH} <10^{ -2.66}$ for $f_{\rm ann} \langle \sigma v \rangle /m_{\chi} = 3 \times 10^{-28} \rm  cm^3 /s/ GeV$. 
The constraint depends on the annihilation rate and, for instance, a weaker bound $f_{\rm PBH} <10^{ -2.38}$ arises for a smaller $f_{\rm ann} \langle \sigma v \rangle /m_{\chi} = 3 \times 10^{-29} \rm cm^3 /s/ GeV$.
The energy injection from DM annihilation can also create the CMB spectral distortion, and we found that the PBH abundance which can generate the CMB distortion observable by the PIXIE-like future observation has been already ruled out by the Planck data. 
The redshifted 21-cm observation is also a promising probe on the heating sources such as the DM annihilation in the dark ages.
The observational determination of the redshift at which the global 21-cm signal shifts from the absorption to the emission can provide the constraint on the efficiency of the heating source in the dark ages because, at this transition redshift, the heated baryon temperature becomes larger than the CMB temperature.
We discussed how this transition redshift is affected in the presence of the PBH isocurvature perturbations.
Recent EDGES observation reported that they detected the absorption signals of global redshifted 21-cm lines between $15 \lesssim z \lesssim 22$.
While the justification of their large absorption signal amplitude is under an active debate, we solely focused on the redshift dependence of the signal and obtained the PBH abundance constraint $ f_{\rm PBH} < 10^{ -3.4}$ for $f_{\rm ann} \langle \sigma v\rangle/m_\chi = 3 \times 10^{-28} {\rm cm^3/s/GeV}$.  We also argued that our bounds on $f_{\rm PBH}$ are insensitive to $M_{\rm PBH}$ for the parameter range of interest in our analysis ($M_{\rm PBH}\gtrsim 10^{-6}M_{\odot}$ when the minimum dark matter halo mass $M_{\rm min}=10^{-6}M_{\odot}$).

Our Poisson effect constraints would be of particular interest for the light (sub-GeV) WIMP which has been less explored compared with the heavier ($>1$ GeV) WIMP in the presence of PBHs.  
For instance, the CMB can give the bounds $f_{\rm PBH}\lesssim {\cal O}(10^{-3})$ for $m_{\chi}=1$ MeV and $f_{\rm ann}\langle \sigma v \rangle =3 \times 10^{-31} \rm cm^3/s$. Our bounds are independent from and complementary to the seed effects which consider the DM accretion into individual PBHs. The DM accretion into PBHs can form the steep profile $\rho(r) \propto r^{-9/4}$ when the DM kinetic energy is negligible compared with their potential energy, and the consequent enhanced DM annihilation can lead to the tight bounds on PBHs using the observation data such as the Fermi gamma ray and Planck CMB data \cite{Boucenna:2017ghj,Adamek:2019gns,Eroshenko:2016yve,Carr:2020mqm,Cai:2020fnq,Delos:2018ueo,Kohri:2014lza,Bertone:2019vsk,Carr:2018rid,1984MNRAS.206..801C,Hertzberg:2020kpm,Kashlinsky:2016sdv,Tashiro:2021xnj}. Even though the detailed numerical analysis has not been performed yet for such seed effects when the dark matter is light such that the kinetic energy cannot be ignored in estimating the dark matter profile around a PBH, the analytical estimation indicates the bounds on $f_{\rm PBH}$ would be significantly weakened for small $m_{\chi}$ and $M_{\rm PBH}$ due to a less steep profile around a PBH \cite{Eroshenko:2019pxt,Boucenna:2017ghj,Carr:2020mqm}. Ref. \cite{Carr:2020mqm}, for instance, analytically estimated the upper bound on $f_{\rm PBH}$ assuming the Maxwell-Boltzmann distribution for the DM velocity and, for example, $f_{\rm PBH}\lesssim 0.1$ for $m_{\chi}\sim 1$ MeV and $M_{\rm PBH}\sim 0.01 M_{\odot}$ (the upper bound of $f_{\rm PBH}$ scales as $\propto m_{\chi}^{-3.7} M_{\rm PBH}^{-1.5}$ assuming the DM kinetic decoupling dependence on the DM mass as $T_{\rm KD}\propto m_{\chi}^{5/4}$ typical for a bino-like WIMP \cite{Bringmann:2006mu}). The bounds on $f_{\rm PBH}$ when the dark matter kinetic energy is not negligible would require a more detailed numerical analysis, and such numerical studies which also should take account of a DM model dependence such as the nature of DM kinetic decoupling \cite{Loeb:2005pm,Bertschinger:2006nq,Gondolo:2012vh, Profumo:2006bv,Gondolo:2016mrz,Green:2003un,Green:2005fa,Bringmann:2006mu} would further quantitatively clarify the complementary between the Poisson and seed effects of PBHs in the presence of the WIMPs.
The bounds on the PBH parameters from DM annihilation when a halo includes multiple PBHs (which is likely when $f_{\rm PBH}>10^{-4}$ \cite{Inman:2019wvr}) and when a PBH interacts with another (e.g. forming the binary) would be also worth pursuing.
\\
\\
This work was supported by the Institute for Basic Science (IBS-R018-D1) and Grants-in-Aid for Scientific Research from JSPS (21K03533). KK thanks the Kobayashi-Maskawa Institute at Nagoya University for the hospitality.

\bibliography{./kenjireference}



\end{document}